\documentclass[preprint,showpacs,preprintnumbers,amsmath,amssymb]{revtex4}
\usepackage{amsfonts}
\usepackage{amssymb}
\usepackage{latexsym}
\usepackage{graphicx}% Include figure files
\usepackage{psfig}
\newcommand{\al}{\alpha}

\newcommand{\be}{\beta}

\newcommand{\si}{\sigma}

\newcommand{\de}{\delta}

\newcommand{\rar}{\rightarrow}

\newcommand{\non}{\nonumber}
\begin{document}

\begin{flushright}
Preprint ICN-UNAM 04-07\\
June 21, 2004
\end{flushright}

\title{The exotic $H_3^{2+}$ ion in a strong magnetic field. Linear
  configuration.}

\author{A.~V.~Turbiner}
 \altaffiliation[]{On leave of absence from the Institute for Theoretical
 and Experimental Physics, Moscow 117259, Russia}
\email{turbiner@nuclecu.unam.mx}
%\affiliation{%
%Instituto de Ciencias Nucleares, Universidad Nacional Aut\'onoma
%de M\'exico, Apartado Postal 70-543, 04510 M\'exico, D.F., Mexico}
%
\author{J.~C.~L\'opez Vieyra}
\email{vieyra@nuclecu.unam.mx}
%\affiliation{%
%Instituto de Ciencias Nucleares, Universidad Nacional Aut\'onoma
%de M\'exico, Apartado Postal 70-543, 04510 M\'exico, D.F., Mexico}

\author{N.~L.~Guevara}
\email{nicolais@nuclecu.unam.mx}
%\affiliation{%
%Departamento de Fisica, Universidad Aut\'onoma Metropolitana,
%09340 M\'exico, D.F., M\'exico.}
\affiliation{%
Instituto de Ciencias Nucleares, Universidad Nacional Aut\'onoma
de M\'exico, Apartado Postal 70-543, 04510 M\'exico, D.F., Mexico}

%\date{March 20, 2004}

\begin{abstract}
  An accurate study of the lowest $1\si_g$ and the low-lying excited
  $1\si_u$, $1\pi_{u,g}$, $1\de_{g,u}$ electronic states of the exotic
  molecular ion $H_3^{2+}$ in linear configuration parallel to a
  magnetic field is carried out.  The magnetic field ranges from
  $10^{10}$\,G up to $4.414 \times 10^{13}$\,G where non-relativistic
  considerations are justified.  The variational method is exploited
  and the {\it same} trial function is used for different magnetic
  fields.  It is shown that the states of positive $z$-parity $1\si_g,
  1\pi_u, 1\de_{g}$ are bound states of the $H_3^{2+}$ exotic ion for
  all magnetic fields studied.  We also demonstrate that for magnetic
  fields $B\gtrsim 2.35\times 10^{12}$\,G the potential energy surface
  well corresponding to the $1\si_g$ state contains at least one
  longitudinal vibrational state. It is also shown that the negative
  $z$-parity states $1\si_u, 1\pi_g, 1\de_{u}$, are purely repulsive
  in the whole range of magnetic fields studied, $B=10^{10}- 4.414
  \times 10^{13}$\,G.

\end{abstract}


\maketitle

\section{Introduction}

Recently, it was announced that in a sufficiently strong magnetic
field $B \gtrsim 10^{11}\,G$ three protons situated on the magnetic
line can be bound by one electron forming the exotic molecular ion
$H_3^{2+}$ in linear configuration \cite{Turbiner:1999}. A major
feature of this configuration is that with a magnetic field growth the
system becomes more and more bound (binding energy grows), and more
and more compact (equilibrium distance decreases).  Later it was shown
that this exotic ion $H_3^{2+}$ in linear configuration becomes even
the most stable one-electron system with smallest total energy for
magnetic fields $B \gtrsim 10^{13}\,G$ \cite{Lopez-Tur:2000}. A recent
study has demonstrated that the $H_3^{2+}$ molecular ion can also
exist in a certain spatial configuration -- the protons form an
equilateral triangle while a magnetic field is directed perpendicular
to it \cite{Turbiner:2002}. The goal of this article is to make a
detailed quantitative study of the ground state and the lowest excited
states of the $H_3^{2+}$-ion in a linear configuration for $B=10^{10}
- 4.414 \times 10^{13}\,G$, which was announced in
\cite{Turbiner:1999}. In particular, this is the first study of the
lowest excited states of the $H_3^{2+}$ ion.

Atomic units are used throughout ($\hbar$=$m_e$=$e$=1), although
energies are expressed in Rydbergs (Ry). The magnetic field $B$ is
given in a.u. with $B_0= 2.35 \times 10^9\,G$.

\section{Generalities}

\begin{figure}[tb]
\begin{center}
   \includegraphics*[width=3in,angle=-90]{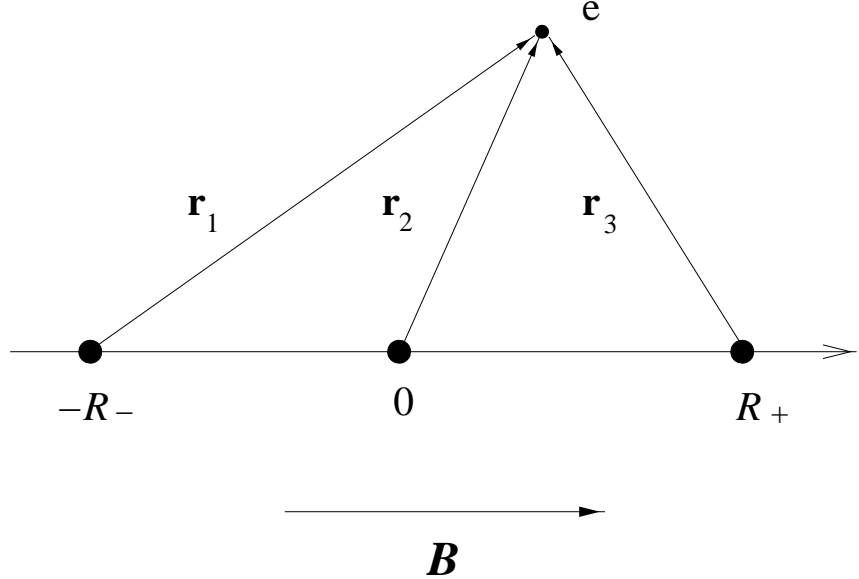}
    \caption{\label{fig:h3pp} Geometrical setting  for the $H_3^{2+}$ 
      ion placed in a magnetic field directed along the $z$-axis. The
      protons (marked by bullets) are situated on the $z$-line at
      distances $R_{\pm}$ from the central proton which is placed at
      the origin.}
\end{center}
\end{figure}

We study the system of three protons and one electron $(pppe)$ placed in a
uniform constant magnetic field.  It is assumed that the protons are
infinitely heavy (Born-Oppenheimer approximation of zero order) and
that they are situated along the magnetic field direction forming a
linear chain.  The Hamiltonian which describes this system when the
magnetic field is directed along the $z-$axis, ${\bf B}=(0,0,B)$ is
written as
\begin{equation}
\label{Ham} {\cal H} = {\hat p}^2 + \frac{2}{R_+} + \frac{2}{R_-}
+ \frac{2}{R_++R_-} -\frac{2}{r_1} -\frac{2}{r_2}
 -\frac{2}{r_3}
  + ({\hat p} {\cal A}+{\cal A}{\hat p}) +  {\cal A}^2 \ ,
\end{equation}
(see Fig.~\ref{fig:h3pp} for the geometrical setting and notations),
where ${\hat p}=-i \nabla$ is the momentum, ${\cal A}$ is a vector
potential which corresponds to the magnetic field $\bf B$. It is
chosen in the symmetric gauge,
\[
   {\cal A}= \frac{B}{2} (-y,\ x,\ 0)\ .
\]
Hence, the total energy $E_T $ of the $H_3^{2+}$-ion is defined as the
total electronic energy plus the Coulomb energy of proton repulsion.
The binding energy is defined as the affinity of the system to form a
bound state with respect to the system when the electron and the three
protons are infinitely separated, $E_b=B-E_T$. There are two
dissociation processes: one of them has in the final state a hydrogen
atom, $H_3^{2+} \rar H + 2p$ while the other one has a $H_2^+$
molecular ion, $H_3^{2+} \rar H_2^+ + p$. Therefore, the first
dissociation energy is defined as the affinity of the system to form a
bound state having two protons at infinity,
$E_{d_{\texttt{atom}}}=E_H-E_T\,$, where $E_H$ is the total energy of
the hydrogen atom in a magnetic field $B$. While the second
dissociation energy is defined as an affinity to form a bound state
having one proton at infinity, $E_{d_{\texttt{ion}}}=E_{H_2^+}-E_T\,$,
where $E_{H_2^+}$ is the total energy of the hydrogen molecular ion
$H_2^+$ in a magnetic field $B$.  A contribution coming from the
spin degrees of freedom is neglected.

The problem we study is characterized by two integrals of motion: (i)
the operator of the $z$-component of the angular momentum (projection
of the angular momentum on the magnetic field direction) giving rise
to the magnetic quantum number $m$ and (ii) the spatial parity
operator $P({\vec r} \rar -{\vec r})$ which has eigenvalue $p=\pm 1$.
Hence, any eigenstate has two explicit quantum numbers assigned: the
magnetic quantum number $m$ and the parity $p$.  Therefore the space
of eigenstates is split into subspaces (sectors) with each of them
characterized by definite values of $m$ and $p$.  It is worth to note
that the Hamiltonian (\ref{Ham}) is also invariant with respect to $z
\to -z$. Therefore, in general one can classify the eigenstates using
the quantum number $\sigma=\pm 1$ for positive/negative $z$-parity
instead of $p$.  However, this classification is related with the
above described - there exists a relation between the quantum numbers
corresponding to the $z$-parity and the spatial parity: $$p=\sigma
(-1)^m\ .$$
To classify eigenstates we follow the convention widely
accepted in quantum chemistry using the quantum numbers $m$ and $p$.
In particular, the notation for the states that we are going to use is
similar to that introduced for $H_2^+$-ion in parallel configuration
\cite{TurbLopez:2004} and is based on the following: the first number
corresponds to the number of excitation - "principal quantum number",
e.g. the number 1 is assigned to the ground state (lowest state), then
Greek letters $\si, \pi, \de$ correspond to the states with
$m=0,-1,-2$, respectively, with subscript $g/u$ (gerade/ungerade)
describing positive/negative parity $p$.

The excited states which we plan to study are the lowest states (of
the type of the ground state) of the sectors with different magnetic
quantum numbers $m$ and $p$. It is quite obvious from a physical point
of view that the ground state of a sector with $m>0$ always has larger
total energy than those with $m \leq 0$. For this reason we restrict
our consideration to the states with $m=0,-1,-2$.

Conservation of the $z$-component of the angular momentum implies that
the electronic wave function (in cylindrical coordinates
$(\rho,\varphi,z)$) can be written as
\begin{equation}
\label{Psi}
\Psi = \,e^{i m \varphi}\rho^{|m|}\,\psi_m \ ,
\end{equation}
where $m$ is the magnetic quantum number. If we gauge rotate the
Hamiltonian (\ref{Ham}) with a factor from (\ref{Psi}), it takes the
form
\begin{equation}
\label{Ham_m}
 {\cal H}_m\ =\
 e^{-i m \varphi}\rho^{-|m|}\, {\cal H} e^{i m \varphi}\rho^{|m|}\
 = {\hat p_m}^2 + \frac{2}{R_+} + \frac{2}{R_-}
+ \frac{2}{R_++R_-} -\frac{2}{r_1} -\frac{2}{r_2} -\frac{2}{r_3}
  + m B +  \frac{B^2 \rho^2}{4} \ ,
\end{equation}
where
\[
 {\hat p_m} = e^{-i m \varphi}\rho^{-|m|}\ {\hat p}\ e^{i m
 \varphi}\rho^{|m|}\ ,
\]
is the gauge rotated momentum (covariant momentum) and $\psi_m$ are
the eigenfunctions. The constant term $mB$ describes the linear Zeeman
effect splitting. In principle this term can be absorbed into the definition
of the total energy.  The representation (\ref{Ham_m}) is rather
convenient since each Hamiltonian for fixed $m$ describes the family
of eigenstates with quantum number $m$ and can be treated
independently of the states with $m'$ different from $m$. Now the
Hamiltonian (\ref{Ham_m}) has only the invariance corresponding to the
spatial parity conservation.

As a method to explore the problem we use the variational procedure.
The recipe of choice of trial functions is based on physical arguments
and is described in full generality in \cite{Tur}, where we address
the reader. The ground state
trial function for fixed $m$ and $p$ is chosen in the form
\begin{equation}
\label{ansatz}
  \psi_m^{(trial)} = A_1\psi_1 + A_2 \psi_2 + A_3\psi_3+
  A_4\psi_4 + A_5 \psi_5 + A_6\psi_6\, ,
\end{equation}
with $\psi_1 \ldots \psi_6$  given by
\begin{subequations}
\label{psi123}
\begin{eqnarray}
%%%%%%%%%%%%%%%%%%%%%%%%%%%%%%%%%%%%%%%%%%%%%%%%%%%%%%%%%%%%
\psi_1   &=& \left\{\begin{array}{ll}
       {e}^{-\al_1 (r_1+r_2+r_3)} {e}^{-B\be_{1}\rho^2/4 }
\, , & \qquad \mbox{if}\ \sigma=+1\, \\
           0 \, ,                                            
& \qquad \mbox{if}\ \sigma=-1\,  \\
\end{array}
\right .
  \label{equationa} \\[7pt]
%%%%%%%%%%%%%%%%%%%%%%%%%%%%%%%%%%%%%%%%%%%%%%%%%%%%%%%%%%%%
\psi_2   &=& \left\{\begin{array}{ll} \big({e}^{-\al_2 r_1} +
{e}^{-\al_2 r_2} +  {e}^{-\al_2 r_3}\big) {e}^{ -  B  \be_{2}
\rho^2/4 } \, , & \qquad \mbox{if}\ \sigma=+1\, \\
           0 \, ,                 & \qquad \mbox{if}\ \sigma=-1\,  \\
\end{array}
\right.   \label{equationb} \\[7pt]
%%%%%%%%%%%%%%%%%%%%%%%%%%%%%%%%%%%%%%%%%%%%%%%%%%%%%%%%%%%%
\psi_3   &=&  \left\{\begin{array}{ll}
\big(  {e}^{-\al_3 (r_1+r_2)} + {e}^{-\al_3 (r_1+r_3)} + 
{e}^{-\al_3 (r_2+r_3)}  \big)
       {e}^{ -  B  \be_{3} \rho^2/4 }
   \, ,  & \qquad \mbox{if}\ \sigma=+1 \, \\
 0 \, ,  & \qquad \mbox{if}\ \sigma=-1 \,  \\
\end{array}
\right.   \label{equationc} \\[7pt]
%%%%%%%%%%%%%%%%%%%%%%%%%%%%%%%%%%%%%%%%%%%%%%%%%%%%%%%%%%%%
\psi_4   &=& \big(
              {e}^{-\al_4 r_1-\al_5 r_2}+
              \sigma {e}^{-\al_5 r_1-\al_4 r_2}+ 
{e}^{-\al_5 r_1-\al_4 r_3}+
              \sigma {e}^{-\al_4 r_1-\al_5 r_3}+ 
\label{equationd} \\&& \hspace{150pt}
              {e}^{-\al_4 r_2-\al_5 r_3}+
              \sigma {e}^{-\al_5 r_2-\al_4 r_3}  \big)
              {e}^{ -  B  \be_{4} \rho^2/4 }\, ,  \non \\[7pt]
%%%%%%%%%%%%%%%%%%%%%%%%%%%%%%%%%%%%%%%%%%%%%%%%%%%%%%%%%%%%
\psi_5   &=& \left\{\begin{array}{ll}
          \big(
                {e}^{-\al_6 (r_1 + r_2) -\al_7 r_3}+
                {e}^{-\al_6 (r_1 + r_3) -\al_7 r_2}+
                {e}^{-\al_6 (r_2 + r_3) -\al_7 r_1} \big)
         {e}^{ -  B  \be_{5} \rho^2/4 } \, , & \quad \mbox{if}
\ \sigma=+1 \, \\
 0 \, ,                 & \quad \mbox{if}\ \sigma=-1 \,  \\
\end{array}
\right. \label{equatione} \\[7pt]
%%%%%%%%%%%%%%%%%%%%%%%%%%%%%%%%%%%%%%%%%%%%%%%%%%%%%%%%%%%%
\psi_6   &=& \big(
          {e}^{-\al_8    r_1 -\al_9    r_2 -\al_{10} r_3 } +
   \sigma {e}^{-\al_9    r_1 -\al_8    r_2 -\al_{10} r_3 } +
          {e}^{-\al_{10} r_1 -\al_8    r_2 -\al_9    r_3 } +  
\label{equationf}\\&&\hspace{50pt}
   \sigma {e}^{-\al_8    r_1 -\al_{10} r_2 -\al_9    r_3 } +
          {e}^{-\al_9    r_1 -\al_{10} r_2 -\al_8    r_3 } +
   \sigma {e}^{-\al_{10} r_1 -\al_9    r_2 -\al_8    r_3 }
\big)
         {e}^{ -  B  \be_{6} \rho^2/4 }\, , \non
\end{eqnarray}
\end{subequations}
where for the sake of convenience of a representation we use the quantum
number $\sigma$. The functions $\psi_1 \ldots \psi_6$ are $S_3$
invariant with respect to the permutations of the identical protons
and $\sigma = \pm 1$ corresponds to the symmetric (antisymmetric) trial
functions.  It is clear that the functions $\psi_1\ldots \psi_6$ are
also eigenfunctions of the $z$-parity operator ${\nobreak (p_z\psi_i =
  \sigma \psi_i, \quad i=1\ldots 6)}$.

In Eqs. (\ref{psi123}), $A_{1,\ldots, 6}$ and $\al_{1,\ldots, 10},
\beta_{1,\ldots,6}$, as well as the internuclear distances $R_{\pm}$
(see Fig.~\ref{fig:h3pp}) are variational parameters \footnote{Due to the
  normalization of the wave function one of the coefficients $A$
  should be kept fixed.  Usually, we put $A_1=1$}.  The total number
of parameters is 23 for the symmetric case ($\sigma=+1$) and 10 for
the antisymmetric case ($\sigma=-1$) \footnote{We always assume
  following symmetry arguments and check it afterwards numerically
  that the optimal configuration corresponds to $R_+=R_-$.
  Effectively it reduces the number of variational parameters.}. The
functions $\psi_{1 \ldots 6}$ carry a certain physical meaning.  The
function $\psi_{1}$ ($\psi_{2}$) describes a coherent (incoherent)
interaction of the electron with the protons. The functions
$\psi_{3},\psi_{4}$ describe a coherent interaction of the electron
with two protons (it may be called as $H_2^+$ type interaction).  In a
certain sense it corresponds to the interaction of $H_2^+$ with a
proton. The function $\psi_{5}$ represents an interaction of the
electron with all three charged centers of the type $(H_2^+ -
H\mbox{-atom})$-mixture. Finally the function $\psi_{6}$ is a
nonlinear interpolation of the functions $\psi_1 \ldots \psi_6$.

Calculations were performed using the minimization package MINUIT from
CERN-LIB. Numerical integrations were carried out with a relative
accuracy of $\sim 10^{-11}$ by use of the adaptive D01FCF routine from
NAG-LIB. All calculations were performed on a dual PC with two Xeon
processors of $2.8$ GHz each. Every particular calculation of a given
eigenstate at fixed magnetic field including minimization has taken
several hours of CPU time. However, when the variational parameters are
found it takes a few seconds of CPU time to calculate the variational
energy.

It is necessary to mention two technical difficulties we encountered.
Calculation of two-dimensional integrals with high accuracy which
appeared in the problem has required a development of a very
sophisticated numerical technique. It was created a `dynamical
partitioning' of the domain of integration, which depends on the
values of the variational parameters (see e.g. \cite{TurbLopez:2003}).
The domain partitioning was changed with a change of the parameters.
Sometimes the number of sub-domains in particular integration was
around 20. Another technical problem is related with a very
complicated profile of the variational energy as function of the
variational parameters which is characterized by many local minima,
saddle points and valleys.  Localization of the global minimum
numerically of such a complicated function with high accuracy is a
difficult technical problem which becomes even more difficult in the
case of twenty or more variational parameters. Examining the physical
relevance of trial functions allows us to avoid spurious minima. The
parameters obtained in (\ref{ansatz}) at every step of minimization
were always examined from the physical point of view.  Such
considerations are always something of an art.

\section{Results}

\subsection{$m=0$ }

The $m=0$ subspace consists of two subspaces, $p=1$ (even states)
and $p=-1$ (odd states).

\subsubsection{$1\si_g$ state ($p = 1$)}

The state $1\si_g$ is the global ground state of the exotic
$H_3^{2+}$-ion.  Its eigenfunction has no nodes (Perron theorem). For
this state our variational trial function (\ref{ansatz}) with
${\nobreak m=0, p=1}$ depends effectively on twenty-two parameters
(see footnote [13]).  In comparison with the trial function used
in the first studies \cite{Turbiner:1999, Lopez-Tur:2000} we have
added two extra ansatz, $\psi_4$ and $\psi_5$. As it was mentioned
above, the search for the global minimum numerically with high
accuracy in the case of so many variational parameters is a difficult
technical task.  We use a sophisticated strategy for localizing the
minimum. As a first step we minimize ansatz by ansatz, then we take a
superposition of two ansatz, then three ansatz etc. An essential
element of the strategy is to impose a (natural) condition that the
variational parameters change smoothly as a function of the magnetic
field $B$. The above-mentioned strategy allowed us to improve our
previous results for the ground state reported in
\cite{Lopez-Tur:2000} on total and binding energies, and also on the
lowest vibrational energies. The qualitative results remain unchanged
(see below).

The performed variational calculations indicate clearly to the
existence of a minimum in the potential energy surface $E_T(R_+,R_-)$
of the $(pppe)$ system for magnetic fields ranging $B=10^{10} -
4.414\times 10^{13}$\,G.  The minimum in the total energy always
corresponds to the situation when $R_+=R_- \equiv R_{eq}^{H_3^{2+}}$
confirming the qualitative symmetry arguments.  Table \ref{T1sg} shows
the results for the total $E_T$ and binding energies $E_b$, as well as
the internuclear equilibrium distance $R_{eq}^{H_3^{2+}}$ for the
ground state $1\si_g$ calculated with the trial function
(\ref{ansatz}) and with the trial function used in our previous
studies (Eq.(\ref{ansatz}) with $A_4=A_5=0$) \cite{Turbiner:1999,
  Lopez-Tur:2000}.  Thus, we predict that the exotic $H_3^{2+}$ ion
can exist even in a larger domain of magnetic field strengths as
reported in \cite{Turbiner:1999, Lopez-Tur:2000}.  We confirm a
general statement that with an increase of the magnetic field strength
the total energies increase, the system becomes more bound (binding
energies increase) and more compact (the internuclear equilibrium
distance decreases).

In general, for all one-electron systems (in linear configuration) the
binding energy increases asymptotically as $\propto \log^2 B$ for
strong magnetic fields (for a discussion about the case of the
hydrogen atom see e.g.  \cite{LL}).  In the domain we study
($B=10^{10}-4.414\times 10^{13}$\,G) the rate of increase of binding
energy for $H_3^{2+}$ is slightly larger than the corresponding rates
for the $H$ atom and for the $H_2^+$ ion (see
\cite{TurbLopez:2003,TurbLopez:2004}).  As a result, for magnetic
fields $B\gtrsim 4\times 10^{10}$\,G the $H_3^{2+}$ ion becomes more
bound than the $H$ atom, and for magnetic fields $B\gtrsim 3\times
10^{13}$\,G, $H_3^{2+}$ becomes the most bound (having the lowest
total energy and the largest binding energy) among the one-electron
systems $H$, $H_2^+$, $H_3^{2+}$.  

An straightforward analysis of the internuclear equilibrium distances
shows that in the domain $B=10^{10}-4.414\times 10^{13}$\,G the rate
of decrease of the internuclear equilibrium distance for the
$H_3^{2+}$ ion is also larger than the corresponding rate for the
$H_2^+$ ion ($R_{eq}^{H_3^{2+}}$ decreases in $\sim 20$ times, while
$R_{eq}^{H_2^{+}}$ decreases in $\sim 10$ times when we go from
$B=10^{10}$\,G to $4.414\times 10^{13}$\,G). Making a comparison
between $H_3^{2+}$ and $H_2^{+}$ one can see that the internuclear
equilibrium distance for $H_3^{2+}$ is always larger than that for
$H_2^+$.  The internuclear distances converge to each other as the
magnetic field increases: $R_{eq}^{H_3^{2+}}=0.110$\, a.u., while
$R_{eq}^{H_2^{+}}=0.102$\, a.u. at $B=4.414\times 10^{13}$\, G (see
\cite{TurbLopez:2003,TurbLopez:2004}).

Another important feature of the system is the behavior of its
dissociation as a function of the magnetic field. There are two
dissociation processes: $H_3^{2+} \to H_2^{+} + p$ (i) and $H_3^{2+}
\to H + 2p$ (ii).  Table~\ref{T1sg} shows the dissociation energies
corresponding to these processes
$\nobreak{E_{d_{\texttt{ion}}}=E_{H_2^+}-E_T\,}$ and
$\nobreak{E_{d_{\texttt{atom}}}=E_H-E_T\,}$, respectively, for
different magnetic fields.  Let us consider the first process (i). A
negative dissociation energy $E_{d_{\texttt{ion}}}$ indicates that the
ion $H_3^{2+}$ is unstable towards the decay $\nobreak{H_3^{2+} \to
  H_2^{+} + p}$.  In particular, for $B\gtrsim 10^{10}$\,G
$E_{d_{\texttt{ion}}}$ is negative and decreases with a magnetic field
growth reaching the minimum at $B\sim 10000$\, a.u.  Then, for further
magnetic field increase the dissociation energy (which is still
negative) starts to increase monotonously. Eventually, for a magnetic
field $B\simeq 3\times 10^{13}$\,G the dissociation energy becomes
zero which means $E_T=E_{H_2^+}$ and then starts to be positive. It
implies that the ion $H_3^{2+}$ becomes more bound than $H_2^{+}$ and,
in fact, this ion becomes the most stable one electron system.  The
above described behavior for the dissociation $H_3^{2+} \to H_2^{+} +
p$ also indicates that, for a broad domain of magnetic fields, there
exist two different values of the magnetic field for which we have the
same dissociation energy $E_{d_{\texttt{ion}}}$.

The situation is different for the second dissociation process (ii),
$H_3^{2+} \to H + 2p$.  The dissociation energy
$\nobreak{E_{d_{\texttt{atom}}}}$ increases monotonously in all the
range of studied magnetic fields.  In the domain $10^{10}-4\times
10^{10}$\,G the dissociation energy $\nobreak{E_{d_{\texttt{atom}}}}$
is negative and it indicates that the ion $H_3^{2+}$ is unstable
towards the decay $H_3^{2+}\to H + 2\,p$. For magnetic fields
$B\gtrsim 4\times 10^{10}$\,G the dissociation energy is positive and
the ion $H_3^{2+}$ is more bound than the $H$ atom.  For comparison
the dissociation energies $E_{d_{{\texttt{atom-ion}}}}$ for the
process $H_2^+ \to H + p$ are also shown in Table~\ref{T1sg}
\footnote{the total energies for the Hydrogen atom were calculated
  using the 7-parametric variational trial function used in
  \cite{TurbPot:2001} but with $B_0=2.35\times 10^{9}$\,G}.  For
magnetic fields $B\gtrsim 10^{11}$\,G, the dissociation energy
$E_{d_{\texttt{atom}}}$ for the dissociation $H_3^{2+} \to H + 2p$ is
smaller than the corresponding dissociation energy for $H_2^+ \to H +
p$ except for the domain $B>10000$\,a.u.

The improvement in the total energy obtained using the trial function
(\ref{ansatz}) in comparison to the results based on the reduced trial
function (\ref{ansatz}) ($A_4=A_5=0$)
\cite{TurbLopez:2003,TurbLopez:2004} is the order of $\sim (1 -
5)\times 10^{-3}$\, Ry for the whole range of magnetic fields studied.
It represents a relative improvement of $\sim 0.01\%-0.05\%$ in the
binding energies (see Table \ref{T1sg}).

We should notice that in our previous studies
\cite{Turbiner:1999,Lopez-Tur:2000} a different definition of the
magnetic field atomic unit was used for unit conversion
($B_0=2.3505\times 10^{9}$\, G). It leads to a relative difference in
the total energies of order of $10^{-4}$\,. It should be taken into
account when a comparison of the results in
\cite{Turbiner:1999,Lopez-Tur:2000} and the present results is made.

We show in Fig.~\ref{fig:parms} the behavior of the variational
parameters of the trial function (\ref{ansatz}) as a function of the
magnetic field strength. In general, the behavior of the parameters is
rather smooth and {\it very} slowly-changing, even though the magnetic
field changes by several orders of magnitude. In our opinion such a
behavior of the parameters of our trial function (\ref{ansatz})
reflects the level of adequacy (or, in other words, indicates the
quality) of the trial function. In practice, the parameters can be
approximated by the splines method and then can be used to study
magnetic field strengths other than those presented here.

%\subsubsection{Vibrational Energies and Energy Profiles}
\paragraph{Potential Energy Surfaces,  Vibrational Energies.}

We carried out a detailed accurate study of the electronic potential
surface $E_T(R_+,R_-)$ around the minimum and also along the valleys 
and around the barriers which ensure the existence of bound states.
It allowed us to estimate accurately the (lowest) longitudinal
vibrational energies. 

Let us first proceed to a description of the valleys of potential
energy surfaces. Every potential surface is characterized by two
valleys originated from the minimum. Those valleys are symmetric with
respect to the bisectriz $R_+=R_-$. Therefore it is sufficient to
study a single valley and in further considerations we will be focused
on one of the valleys which is almost horizontal.  In
Fig.~\ref{fig:valleys} the valleys in the $(R_+,R_-)$ plane and the
profile along a valley are shown for different magnetic fields.  Every
profile is characterized by a minimum which corresponds to the
equilibrium ($R_+=R_-=R_{eq}$) and a potential barrier. The height of
the potential barrier is defined with respect to the energy
corresponding to the equilibrium position (minimum) $\Delta E_T =
E_T^{max} - E_T^{min}$. The asymptotics of the profile corresponds to
decay of the system to $H_2^+ + p$, for example, when $R_+\to \infty$
(and $R_-\to R_{eq}^{H_2^+}$) or similarly $R_-\to \infty$ (and
$R_+\to R_{eq}^{H_2^+}$).

The pattern of the valley in the $(R_+,R_-)$ plane exhibits a rather
complicated behavior (see Fig.~\ref{fig:valleys}, where the behavior
near minimum is displayed in amplified form). For example, for the
horizontal valley the $R_-$ firstly grows reaching some maximum, then
decreases reaching a minimum and after that increases approaching to
asymptotics in $R_+$ from below. A similar picture holds for all
magnetic fields studied. We do not think that this type of behavior is
an artifact of insufficient accuracy of our calculations. In
principle, the energy profile curve allows us to estimate the lifetime
of the system with respect to the decay $H_3^{2+}\to H_2^+ + p$ for
magnetic fields $10^{10}\,\mbox{G}\lesssim B \lesssim 4\times
10^{13}$\,G.  However, such study is beyond the scope of the present
analysis.

Table \ref{Tvib} shows the values of the height of the potential
barrier ($\Delta E_T = E_T^{max} - E_T^{min}$) along the valley of
total energy and the position of the top of the energy barrier. From
this table one can see that for the domain of magnetic fields
$B=10^{11}-4.414\times 10^{13}$\,G the height of the potential barrier
increases dramatically from $\sim 0.01$\,Ry up to $\sim 5$\,Ry, 
respectively\footnote{A comparison of the present results for the
  height of the potential barrier, with those reported in
  \cite{Lopez-Tur:2000} shows an increasing loss of accuracy of the
  latter with increasing magnetic field (their ratio varies between
  $\sim 0.2 - 0.8$ in the domain of magnetic fields $10^{11} -
  4.414\times 10^{13}$\,G respectively.}.

Now let us proceed to a calculation of the lowest longitudinal
vibrational state.  The method we use is based on the harmonic
approximation of the potential around the minimum.  All necessary
definitions to perform the analysis are similar to those which are
used to study the vibrational states of linear triatomic molecules and
can be found in standard textbooks \hbox{(see, for
  example,~\cite{goldstein})}.  Following the settings of
Fig.\ref{fig:h3pp}, we find easily the normal coordinates: $R_s =
\frac{1}{\sqrt{2}} (R_+ + R_-)$ (it corresponds to the ``symmetric''
normal mode), and $\nobreak{R_a = \frac{1}{\sqrt{2}} ( R_+ - R_-)}$
(it corresponds to the ``anti-symmetric'' normal mode).  The lowest
vibrational energy is then calculated as
$\nobreak{E_0^{vib}=E_0^{s}+E_0^{a}}$, where $\nobreak{E_0^s=
  \sqrt{\frac{k_s}{m_p}}}$ and $\nobreak{E_0^a=
  \sqrt{\frac{3\,k_a}{m_p}}}$ are the lowest ``symmetric'' and
``anti-symmetric'' vibrational energies respectively, $m_p=1836.15$ is
the proton mass measured in units of the electron mass, and $k_s$,
$k_a$ are the curvatures taken in a.u. near equilibrium for the
symmetric and antisymmetric modes, respectively.

The results for the lowest vibrational energies $E_0^{vib}$ for
different magnetic fields are shown in Table \ref{Tvib}. These results
indicate that with a magnetic field growth the form of the potential
well around the minimum becomes sharper and the lowest vibrational
energies increase drastically growing from $\sim 0.1$\,Ry at
$B=10^{11}$\,G up to $\sim 2$\,Ry at $B=4.414\times
10^{13}$\,G\,\footnote{A comparison of the present results for the
  lowest vibrational energies with our previous results
  \cite{Lopez-Tur:2000} shows that there is a difference in a factor
  $\sim 2$. It appears due to a mistake in normalization of the normal
  modes used in \cite{Lopez-Tur:2000} (cf.Table I therein).}.  Then,
we can conclude that the energy well contains at least one vibrational
state for magnetic fields $B\gtrsim 1000$\,a.u.

\paragraph{Electronic distributions.}
The equilibrium configuration is characterized by a one-peak
electronic distribution centered around the middle proton, being
drastically shrinked in the direction transverse to the magnetic field
as compared with the longitudinal direction (see
Fig.~\ref{fig:edist1}, it is worth to note that the different scales
are used for the $x$ and $z$ axes). This one-peak form of the
electronic distribution for equilibrium position is found for all
magnetic fields studied.  Also the electronic cloud has always a
needle-like form.  When the system is out of equilibrium, the
electronic distributions show the appearance of a pronounced shoulder
which follows the position of the more distant proton, reducing its
height as the system approaches to the decaying configuration, where
the electronic density rearranges to mimic the electronic distribution
of the $H_2^+$ ion.  Nothing especial seems to occur for the
configuration near the top of the potential barrier.  To illustrate
the above-mentioned features we present on Fig.~\ref{fig:edist1} the
normalized electronic density distributions ${\Psi^2(x,y=0,z)}/{\int
  \Psi^2(x,y,z) d{\vec r}}$ and the corresponding contour distribution
for the magnetic field $B=10000$\,a.u. for different positions of the
system $H_3^{2+}$ along the valley in potential surface in $R_+,R_-$.
In this presentation we always keep the central proton at $z=0$. The
following points along the valley are chosen: (a) the equilibrium
position, $R_+=R_-=0.13$\,a.u., (b) a point between positions of
minimum and maximum, $R_+=0.23$\,a.u., (c) a point close to the
maximum, $R_+=0.35$\,a.u.  (maximum is around $R_+\simeq 0.36$\,a.u.
-- see Table~\ref{Tvib}) and (d) a point on the asymptotic part of the
valley where one of the protons is practically detached from the
system.

%%%%%%%%%%%%%  TABLE:1  %%%%%%%%%% 
%%%%%%%%%%%%%  ET, Eb, Req, Dissociation energies 1sigma_g
\begingroup
\squeezetable
\begin{table*}
\caption{\label{T1sg}
  Total $E_T$ and binding $E_b$ energies,  internuclear equilibrium
  distances $R_{eq}$ and dissociation energies   
  $E_{d_{\texttt{ion}}}=E_{H_2^+}-E_T\,$ 
  $(H_3^{2+}\to H_2^+ + p)$ and $E_{d_{\texttt{atom}}}=E_H-E_T\,$ 
  $(H_3^{2+}\to H + 2p)$   for the state $1\si_g$. Dissociation
  energies $E_{d_{{\texttt{atom-ion}}}}=E_H - E_{H_2^+} $ $(H_2^{+}\to
  H + p)$ are shown for comparison. Results
  marked by asterisks  (${}^\star$) correspond to a trial function
  (\ref{ansatz}) with $A_{4,5}=0$ (four Ansatz function, Turbiner and
  L\'opez-Vieyra, unpublished 2003).}
\begin{ruledtabular}
\begin{tabular}{lccccccc}
%    \hline
    $B$  &  $E_T$ (Ry) &  $E_{b}$ (Ry) & $R_{eq}$ (a.u.)
 &$E_{d_{\texttt{ion}}}$ (Ry)& $E_{d_{\texttt{atom}}}$ (Ry)
 & $E_{d_{\texttt{atom-ion}}}$ (Ry) \\
    \hline
%    \hline
$10^{10}$\,G    &     1.8424     &  2.4129     &  2.072
&-0.7519 & -0.2020   & 0.5500 & present      \\
                &     1.8438     &  2.4116     &  2.061  &&  && (${}^\star$) \\
$10$\,a.u.      &     6.6084     &  3.3916     &  1.431
&-0.9581 & -0.1039  & 0.8542 &present
\\
                &     6.6094     &  3.3906     &  1.429   && && (${}^\star$) \\
$10^{11}$\,G    &    36.4297     &  6.1234     &  0.801
&-1.3865 & 0.4105 & 1.7966 &present      \\
                &    36.4327     &  6.1205     &  0.799  &&  && (${}^\star$) \\
$100$\,a.u.     &    91.3611     &  8.6389     &  0.579
&-1.6521 & 1.0596  & 2.7117 &present      \\
                &    91.3655     &  8.6345     &  0.578  &&  && (${}^\star$) \\
$10^{12}$\,G    &   410.3739     & 15.1580     &  0.345
&-1.9845 & 3.2427  & 5.2272  &present      \\
                &   410.3762     & 15.1558     &  0.345  &&  && (${}^\star$) \\
$1000$\,a.u.    &   979.2171     & 20.7829     &  0.259
&-1.9956 & 5.4615   & 7.4571 &present      \\
                &   979.2206     & 20.7794     &  0.259  &&  && (${}^\star$) \\
$10^{13}$\,G    &  4220.9286     & 34.3905     &  0.166
&-1.3634        & 11.8395   & 13.2029  &present      \\
                &  4220.9320     & 34.3872     &  0.166  &&  && (${}^\star$) \\
$10000$\,a.u.   &  9954.5918     & 45.4082     &  0.130
&-0.3890  & 17.4532 &   17.8422  &present      \\
                &  9954.5972     & 45.4028     &  0.130  &&  && (${}^\star$) \\
$4.414\times 10^{13}$\,G
                & 18727.7475     & 55.2312     &  0.110
&0.7294  & 22.7305   & 22.0010   &present      \\
                & 18727.7521     & 55.2267     &  0.110  &&  && (${}^\star$)
%\hline
\end{tabular}
\end{ruledtabular}
\end{table*}
\endgroup

%%%%%%%%%%%%%  TABLE:2 %%%%%%%%%%%
%%%%%%%%%%%%%  vibrational energies and potential barrier 1sigma_g
\begin{table*}
  \caption{\label{Tvib}
    Energies of the lowest longitudinal vibrational
    states $(E_0^{vib})$, height
    of the potential barrier ($\Delta E_T = E_T^{max}-E_T^{min}$)  
    and position of the maximum $(R_+(E_{max}),R_-(E_{max}))$ in 
    the valley corresponding to the
    ground state  $1\si_g$  of the $H_3^{2+}$ molecular ion in a
    magnetic field. Results marked  by asterisks   ${}^{\star}$
    are interpolated values of the height of the barrier using a 
    cubic polynomial fit
    in the variable $\log(1+B^2)$ and  the position of the maximum 
    using a
    fit of the form $R_{\pm}(E_{max},B)=1/P_2^{\pm}$,  where
    $P_2^\pm$ is a quadratic polynomial in the variable 
    $\log(1+B^2)$.}
\begin{ruledtabular}
\begin{tabular}{lcccc}
\( B \) & \( E_0^{vib} \) (Ry)& \( \Delta E_T \)
(Ry)&$R_+(E_{max})$ (a.u.)&$R_-(E_{max})$ (a.u.)\\[-7pt]
               &  &  & &  \\ \hline
%\( 10^{10} \) G
%               & --- & --- &---&--- \\
%% \hline
%\( 10 \) a.u.
%               & 0.0417 & ---  &---&--- \\
% \hline
\( 10^{11} \) G
               & 0.0951 & 0.0101 & 0.94 &0.758 \\
% \hline
\( 100 \) a.u.
               & 0.154 & 0.1007${}^{\star}$ & 0.77 $^{\star}$& 0.517 $^{\star}$  \\
% \hline
\( 10^{12} \) G
               & 0.338 & 0.267 &0.58&0.294  \\
% \hline
\( 1000 \) a.u.
               & 0.530 & 0.581 ${}^{\star}$ &0.50$^{\star}$ &0.221 $^{\star}$  \\
% \hline
\( 10^{13} \) G
               & 1.088 & 1.895 &0.40&0.147  \\
% \hline
\( 10000 \) a.u.
               & 1.594 & 3.339${}^{\star}$ &0.36 ${}^\star$& 0.118 ${}^\star$  \\
%%%%%%         & ---   & 3.337             &0.39           &0.118   \\
% \hline
\( 4.414\times 10^{13} \)G
               & 2.078 & 4.815 &0.33&0.101
% \hline
\end{tabular}
\end{ruledtabular}
\end{table*}

%\clearpage

\subsubsection{$1\si_u$ state ($p=-1$)}

We have performed a detailed study for the state $1\si_u$ of the
$H_3^{2+}$ ion in symmetric configuration $R_+=R_-\equiv R$ in the
domain of magnetic fields $B=10^{10}-4.414\times 10^{13}$\,G.  The
calculations were carried out using the trial function (\ref{ansatz})
with $m=0$ and $p=-1$ corresponding to a state of negative $z$-parity
($\sigma=-1$).  For a broad domain of internuclear distances $R$ the
electronic potential curve $E_T=E_T(R)$ shows no indication of the
existence of a minimum.  Hence, it is natural to conclude that this
state is a purely repulsive state for $B=10^{10}-4.414\times
10^{13}$\,G.  It means that the molecular ion $H_3^{2+}$ does not
exist in the $1\si_u$ state as a bound state.

\subsection{$m=-1$}

The subspace consists of two subspaces, $p=1$ (even states) and
$p=-1$ (odd states).

\subsubsection{$1\pi_u$ state ($p = -1$)}

We study the $1\pi_u$ state of $(pppe)$ system in symmetric
configuration $R_+=R_-\equiv R$.  The trial function (\ref{ansatz}) is
taken with $m=-1$ and ${\nobreak p=-1}$ (which implies $\si=1$).  The
variational calculations indicate that the potential energy curve
$E_T=E_T(R)$ has clear minimum for magnetic fields $B=2.35\times
10^{10}\,{\rm G} - 4.414 \times 10^{13}$\,G. It manifests that
$H_3^{2+}$ ion exists in $1\pi_u$ state. The results are presented in
Table {\ref{T1pu}}.  Similar to the state $1\si_g$ the binding energy
of the $1\pi_u$ state grows steadily with a magnetic field increase
while the equilibrium distance shrinks in a quite drastic manner. For
small magnetic fields the state $1\pi_u$ is more extended than
$1\si_g$.  The equilibrium distance $R_{eq}$ for $1\pi_u$ is much
larger than this distance for the $1\si_g$ state, as for large
magnetic fields these equilibrium distances become comparable.  Among
the $m=-1$ states the state $1\pi_u$ has the smallest total energy.  A
comparison with the corresponding energy of the $1\pi_u$ state of
$H_2^+$ (cf.  Table VII in~\cite{TurbLopez:2004}) shows that the
energy of the $1\pi_u$ state of $H_3^{2+}$ is always larger for
$B=2.35\times 10^{10}\,{\rm G} - 4.414 \times 10^{13}$\,G.

Transition energy from the $1 \sigma_g$ state to the $1\pi_u$ state,
$E_{1 \pi_u}-E_{1 \sigma_g}$ is easily calculated by taking data from
Tables \ref{T1sg} and {\ref{T1pu}}. In the whole range of magnetic
fields studied the transition energy increases monotonically with the
magnetic field growth as expected (see Fig.~\ref{fig:trans}).

In Fig.~\ref{fig:edist2}(a) a plot of the normalized electronic
density distribution ${\Psi^2(x,y=0,z)}/{\int \Psi^2(x,y,z) d{\vec
    r}}$\, for the $1 \pi_u$ excited state in a magnetic field of
$B=10000$\,a.u. is shown. By looking at the corresponding contour
distribution it is evident that the electronic cloud has a torus-like
axially-symmetric form with respect to the $z$-axis.  The radial size
of the torus\footnote{We can defined the radius of the torus-like form
  as the distance from the origin of the $z$-axis to the point of
  maximal probability.}  decreases with a magnetic field increase.  A
similar qualitative behavior of the electronic density distribution is
observed for different magnetic fields .

%%%%%%% TABLE 3 %%%%%%%%%%%%%%%
\begin{table*}
    \caption{\label{T1pu} Total $E_T$, binding $E_b$ energies and 
      equilibrium distance $R_{eq}$
      for the excited state $1\pi_u$.}
\begin{ruledtabular}
\begin{tabular}{lccc}
%    \hline
    $B$  &  $E_T$ (Ry) &  $E_{b}$ (Ry) & $R_{eq}$ (a.u.)   \\
    \hline
%    \hline
%$10^{10}$\,G    &      ---    &   ---      &   ---    \\
$10$\,a.u.      &    7.9289   &   2.0711   &  2.413   \\
$10^{11}$\,G    &   38.6589   &   3.8943   &  1.238   \\
$100$\,a.u.     &   94.3927   &   5.6073   &  0.869   \\
$10^{12}$\,G    &  415.3661   &  10.1658   &  0.497   \\
$1000$\,a.u.    &  985.7952   &  14.2048   &  0.366   \\
$10^{13}$\,G    & 4231.0542   &  24.2649   &  0.226   \\
$10000$\,a.u.   & 9967.3669   &  32.6331   &  0.174   \\
$4.414\times 10^{13}$\,G
                & 18742.7564  &  40.2223   &  0.145   \\
%\hline
\end{tabular}
\end{ruledtabular}
\end{table*}

\subsubsection{$1\pi_g $ state ($p=1$)}

We have performed a detailed study for the state $1\pi_g$ of the
$H_3^{2+}$ ion in symmetric configuration $R_+=R_-\equiv R$ in the
domain of magnetic fields $B=10^{10}-4.414\times 10^{13}$\,G.  The
calculations were carried out using the trial function (\ref{ansatz})
with $m=-1$ and $p=1$ corresponding to a state of negative $z$-parity
($\sigma=-1$).  For a broad domain of internuclear distances $R$ the
electronic potential curve $E_T=E_T(R)$ shows no indication of the
existence of a minimum.  Hence, it is natural to conclude that this
state is a purely repulsive state for $B=10^{10}-4.414\times
10^{13}$\,G.  It means that the molecular ion $H_3^{2+}$ does not
exist in the $1\pi_g$ state as a bound state.

\newpage

\subsection{$m=-2$}

The subspace consists of two subspaces, $p=1$ (even states) and
$p=-1$ (odd states).

\subsubsection{$1\de_g$ state ($p=1$)}

We study the $1\de_g$ state of $(pppe)$ system in symmetric
configuration $R_+=R_-\equiv R$.  The trial function (\ref{ansatz}) is
taken with $m=-2$ and ${\nobreak p=1}$ (which implies $\si=1$).  The
variational calculations indicate that the potential energy curve
$E_T=E_T(R)$ has clear minimum for magnetic fields $B=2.35\times
10^{10}\,{\rm G} - 4.414 \times 10^{13}$\,G. It manifests that
$H_3^{2+}$ ion exists in $1\de_g$ state. The results are presented in
Table {\ref{T1dg}}.  Similar to the states $1\si_g$ and $1\pi_u$ the
binding energy of the $1\de_g$ state grows steadily with a magnetic
field increase while the equilibrium distance shrinks in a quite
drastic manner. For small magnetic fields the equilibrium distance
$R=R_{eq}$ for $1\de_g$ is much larger than this distance for the
$1\si_g$ and $1\pi_u$ state, as for large magnetic fields all three 
equilibrium distances become comparable.  Among the $m=-2$ states the
state $1\de_g$ has the smallest total energy.  A comparison with the
corresponding energy of the $1\de_g$ state of $H_2^+$ (cf. Table IX
in~\cite{TurbLopez:2004}) shows that the energy of the $1\de_g$ state
of $H_3^{2+}$ is always larger for $B=2.35\times 10^{10}\,{\rm G} -
4.414 \times 10^{13}$\,G.

Transition energies from the $1\pi_u$ state to the $1\delta_g$ state
$E_{1 \de_g}-E_{1 \pi_u}$ can be easily calculated from Tables
\ref{T1pu} and \ref{T1dg} showing a monotonically increasing behavior
in all range of magnetic fields studied (see Fig.~\ref{fig:trans}). It
is worth to note that the transition energy from $1\pi_u$ state to the
$1\delta_g$ state is always smaller than the transition energy from
$1\sigma_g$ state to the $1\pi_u$ state. 

In Fig.~\ref{fig:edist2}(b) a
plot of the normalized electronic density distribution
${\Psi^2(x,y=0,z)}/{\int \Psi^2(x,y,z) d{\vec r}}$\, for the $1
\delta_g$ excited state in a magnetic field of $B=10000$\,a.u. is
shown.  Similar to the case of the $1\pi_u$ state, the electronic
cloud for the $1\de_g$ state has a torus-like axially symmetric form
with respect to the magnetic field line ($z$-axis). It is worth to
note that the radial size of this torus-like form of the electronic
distribution for the $1\de_g$ state is larger than for the $1\pi_u$
state.

%%%%% TABLE 4 %%%%%%%%  1\delta_g state %%%%%%%%%%%%%%%%
\begin{table*}
    \caption{\label{T1dg} Total $E_T$, binding $E_b$ energies and 
     equilibrium distance $R_{eq}$ for the state  $1\delta_g$.}
\begin{ruledtabular}
\begin{tabular}{lccc}
%    \hline
    $B$  &  $E_T$ (Ry) &  $E_{b}$ (Ry) & $R_{eq}$ (a.u.)   \\
    \hline
%    \hline
%$10^{10}$\,G    &     ---    &    ---     &   ---   \\
$10$\,a.u.      &    8.3644  &  1.6356    & 3.214   \\
$10^{11}$\,G    &   39.4450  &  3.1082    & 1.548   \\
$100$\,a.u.     &   95.4919  &  4.5081    & 1.070   \\
$10^{12}$\,G    &  417.2496  &  8.2823    & 0.601   \\
$1000$\,a.u.    &  988.3293  & 11.6707    & 0.437   \\
$10^{13}$\,G    & 4235.0826  & 20.2366    & 0.266   \\
$10000$\,a.u.   & 9972.5387  & 27.4613    & 0.202   \\
$4.414\times 10^{13}$\,G
                & 18748.9067 & 34.0720    & 0.167   \\
%\hline
\end{tabular}
\end{ruledtabular}
\end{table*}

%\clearpage

\subsubsection{$1\de_u$ state ($p=-1$)}

We have performed a detailed study for the state $1\de_u$ of the
$H_3^{2+}$ ion in symmetric configuration $R_+=R_-\equiv R$ in the
domain of magnetic fields $B=10^{10}-4.414\times 10^{13}$\,G.  The
calculations were carried out using the trial function (\ref{ansatz})
with $m=-2$ and $p=-1$ corresponding to a state of negative $z$-parity
($\sigma=-1$).  For a broad domain of internuclear distances $R$ the
electronic potential curve $E_T=E_T(R)$ shows no indication to the
existence of a minimum.  Hence, one can conclude that this state is a
purely repulsive state for $B=10^{10}-4.414\times 10^{13}$\,G.  It
means that the molecular ion $H_3^{2+}$ does not exist in the $1\de_u$
state as a bound state.

%% %%%%%%%%%%% Lowest excited states (Total energies)
%% %%%%%%% TABLE 5  %%%%%%%%%%%%%%%%%%
%% %%%%%%%%%%% Lowest excited states (Total energies)
%% %\begingroup
%% %\squeezetable
%% \begin{table*}[hb]
%%     \caption{\label{TETs} Comparison of the total energies $E_T$ (in Rydbergs)
%%       for the low-lying bound states of the $H_3^{2+}$ molecular ion for
%%       magnetic fields $10$\,a.u. - $4.414\times 10^{13}$\,G.}
%%     \begin{ruledtabular}
%%     \begin{tabular}{lccc}
%%  $B$         &$1\sigma_g$& $1\pi_u$ & $1\delta_g$  \\
%% \hline
%% $10$\,a.u.    &     6.6084  &    7.9289  &     8.3644  \\[5pt]
%% $10^{11}$\,G  &    36.4297  &   38.6589  &    39.4450  \\[5pt]
%% $100$\, a.u.  &    91.3611  &   94.3927  &    95.4919  \\[5pt]
%% $10^{12}$\,G  &   410.3739  &  415.3661  &   417.2496  \\[5pt]
%% $1000$\, a.u. &   979.2171  &  985.7952  &   988.3293  \\[5pt]
%% $10^{13}$\, G &  4220.9286  & 4231.0542  &  4235.0826  \\[5pt]
%% $10000$\, a.u.&  9954.5918  & 9967.3669  &  9972.5387  \\[5pt]
%% $4.414\times 10^{13}$\, G
%%               & 18727.7475  & 18742.7564 & 18748.9067  \\[5pt]
%%     \end{tabular}
%% \end{ruledtabular}
%% \end{table*}
%% %\endgroup
%% %\clearpage

%%%%%%%%%%%%%%  FIGURE:1  %%%% 
%%%%%%%%%%%%%%  H3++ geometrical settings

%%%%%%%%%%%%%%%%%  FIGURE:2 %%%%%%%%%%%%%%%%%
%%%%%%%%%%%%%%%%%  Variational parameters as function of B   

% PART A %%%%%%%%%%%%%%%%%%%%%%%%%%%%%%%%
\begin{figure}[h]
  \unitlength=1in
  \begin{center}
  \begin{picture}(6.47699,7.2)(0,0)
%%%%%%%%%%%%%%%%%%%%%%%%%%%%%%%%%%%%%%%%%%%%%%%%%%%%%%%%%%%%%%%%%%%%
  \put(0,0){{\psfig{figure=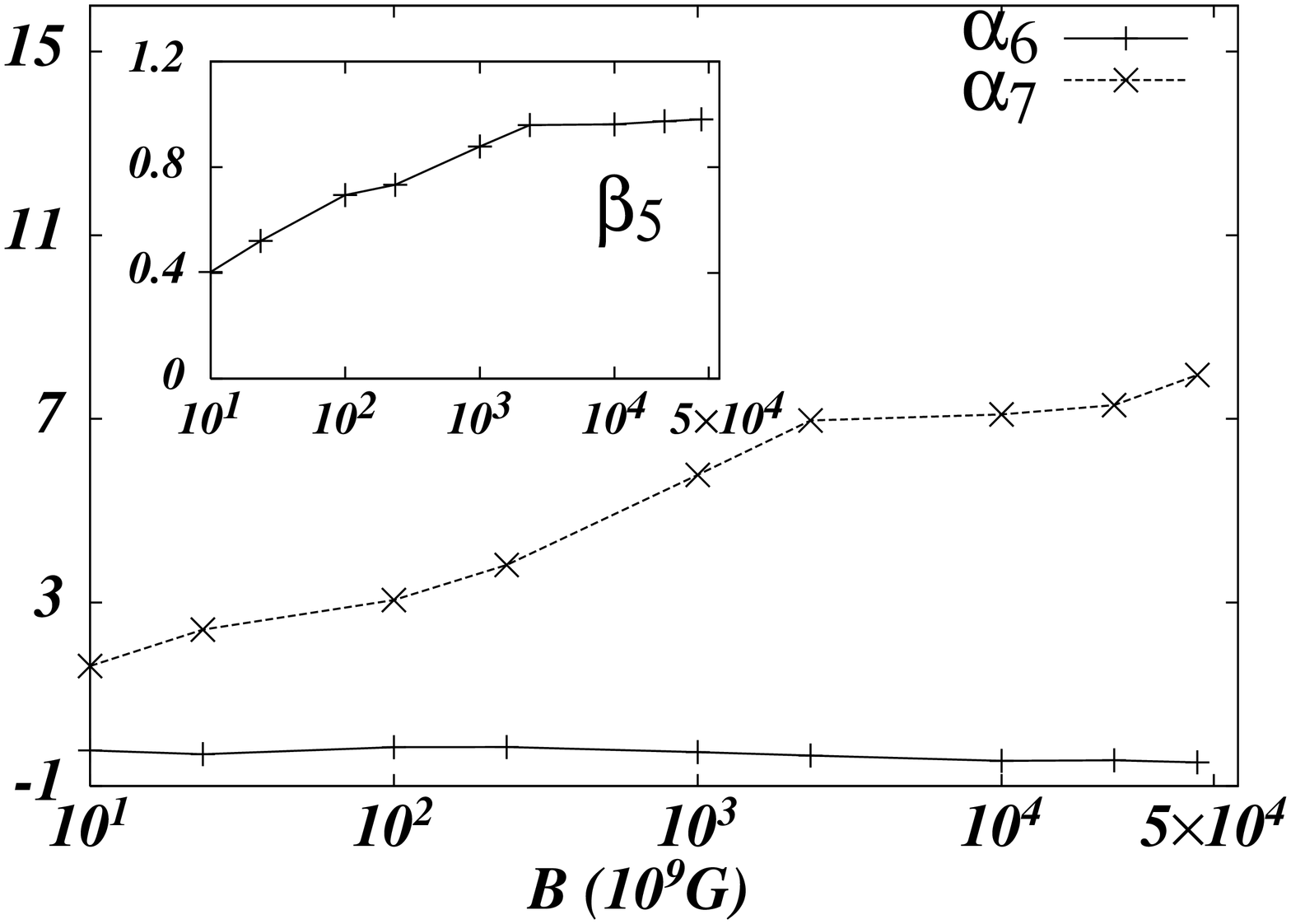,width=3.238in,angle=-90}}}
  \put(3.23,0){{\psfig{figure=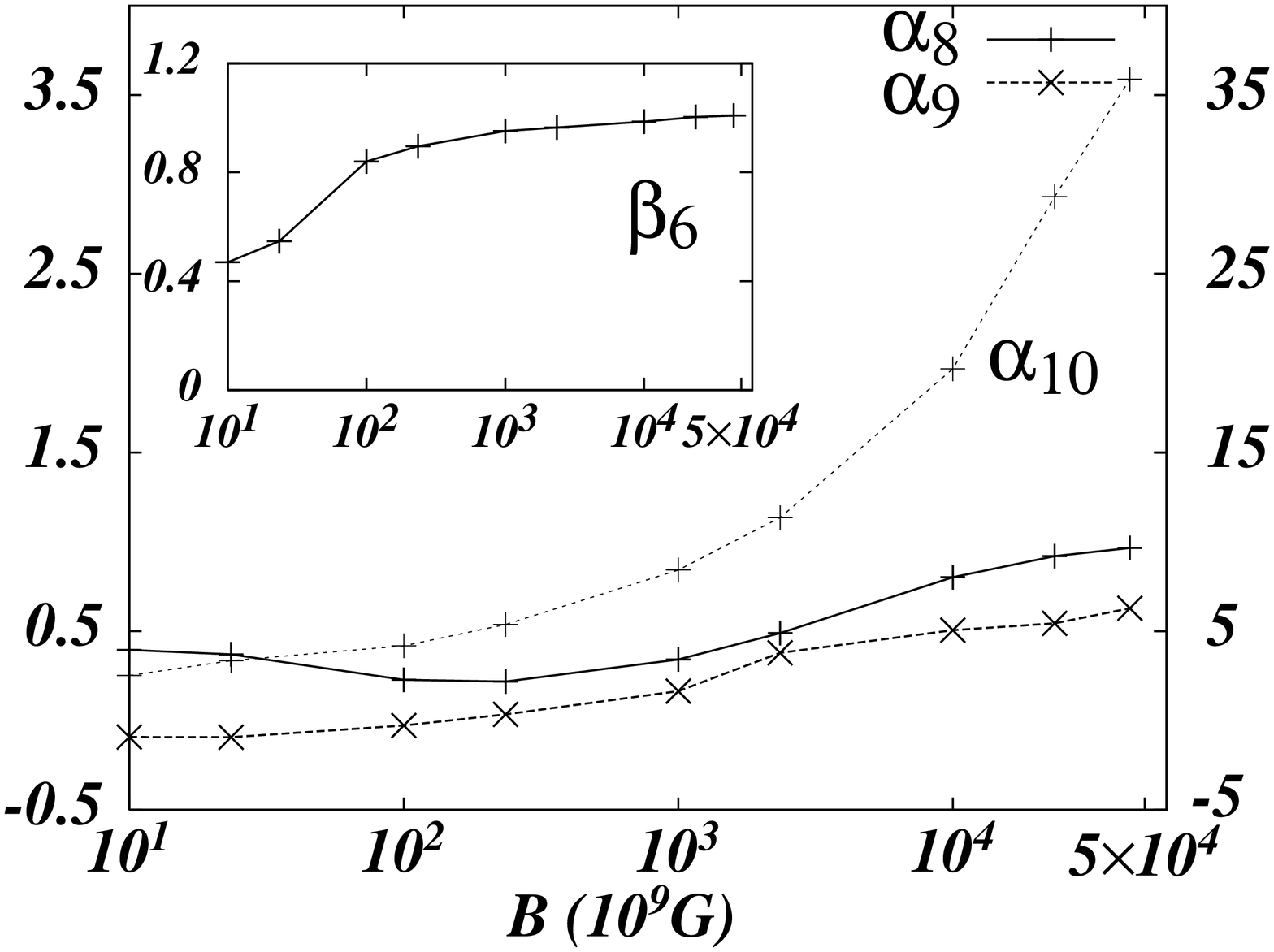,width=3.238in,angle=-90}}}
%%%%%%%%%%%%%%%%%%%%%%%%%%%%%%%%%%%%%%%%%%%%%%%%%%%%%%%%%%%%%%%%%%%%
  \put(0,2.4){{\psfig{figure=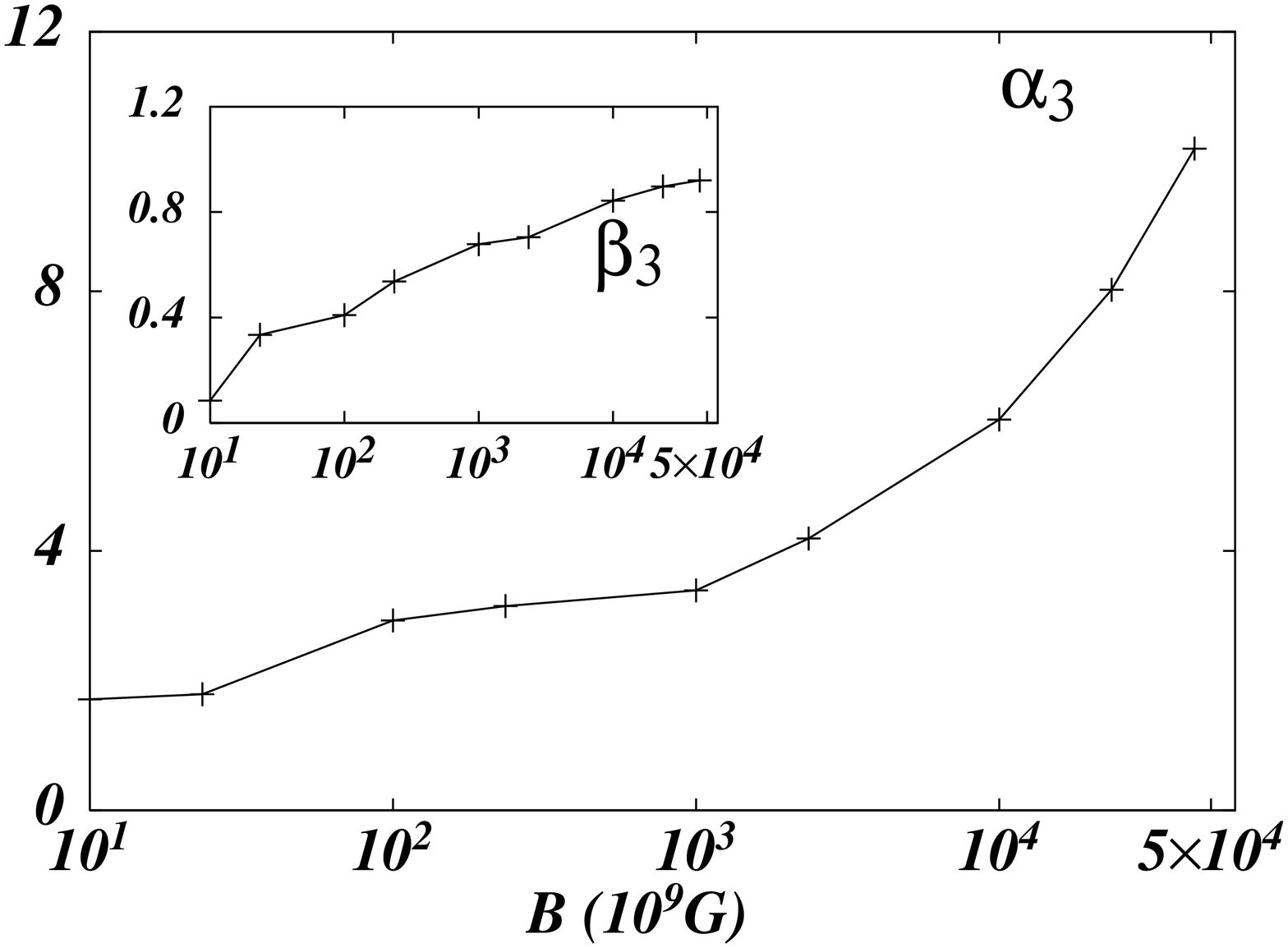,width=3.238in,angle=-90}}}
  \put(3.23,2.4){{\psfig{figure=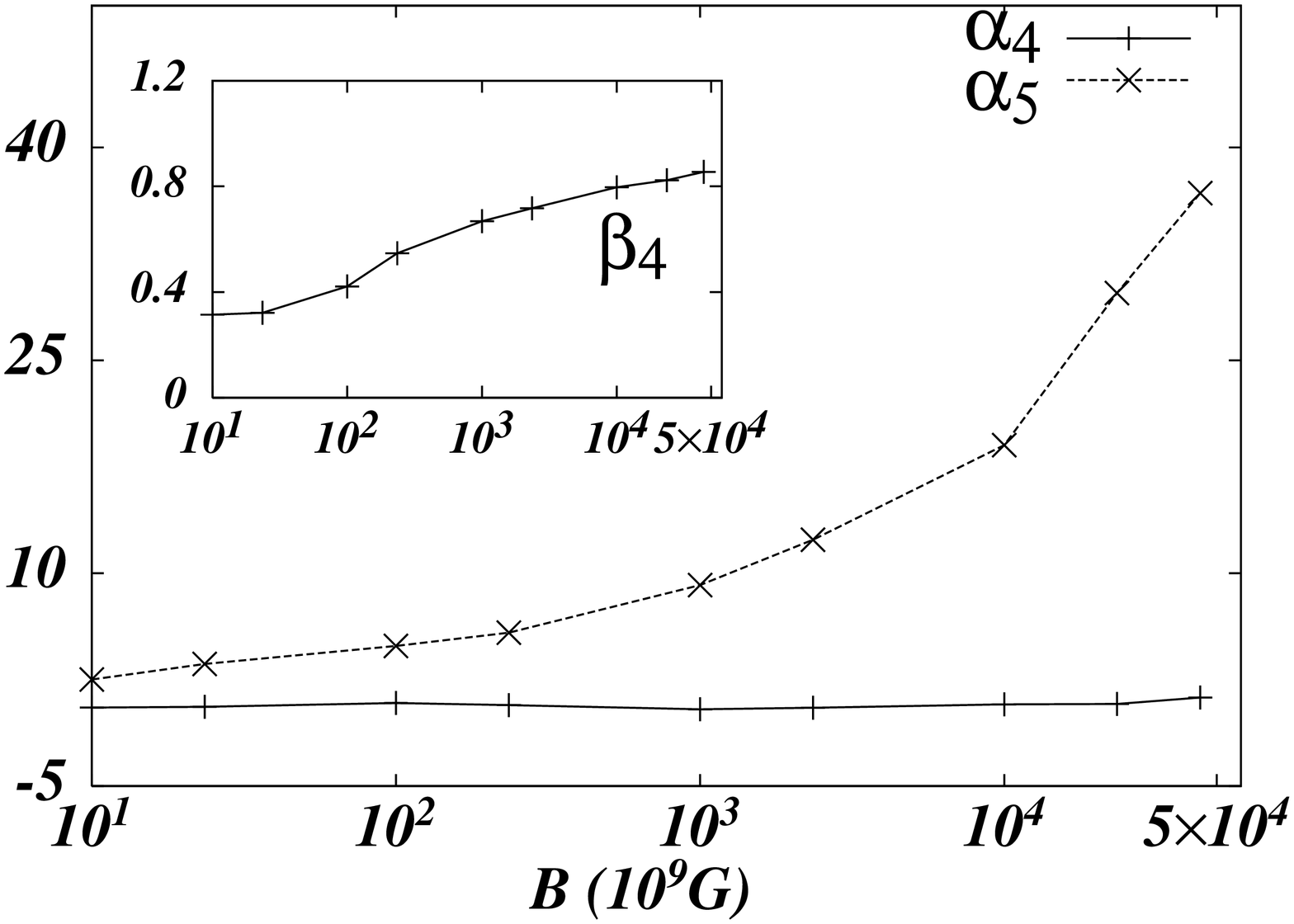,width=3.238in,angle=-90}}}
%%%%%%%%%%%%%%%%%%%%%%%%%%%%%%%%%%%%%%%%%%%%%%%%%%%%%%%%%%%%%%%%%%%%
  \put(0,4.8){{\psfig{figure=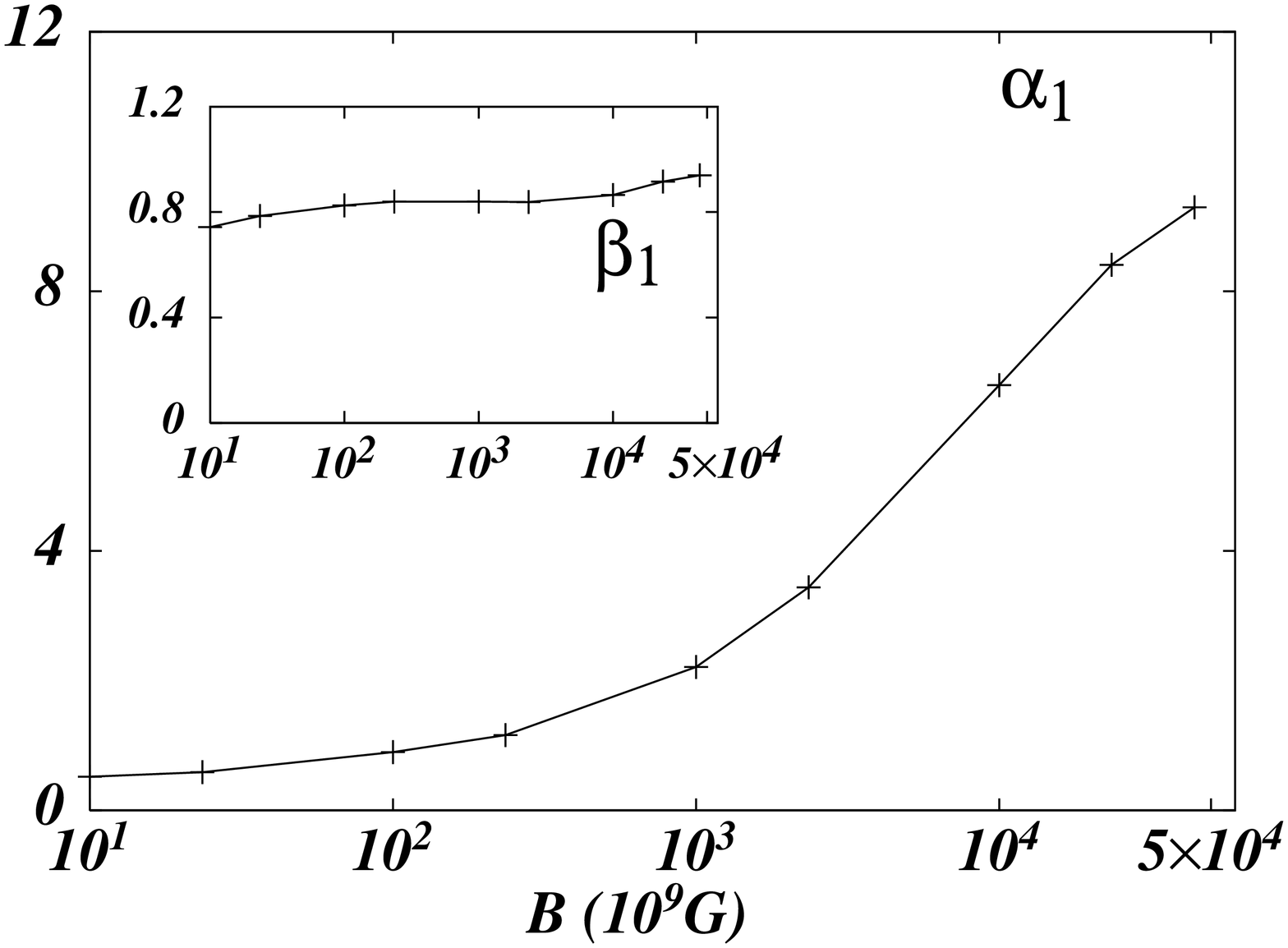,width=3.238in,angle=-90}}}
  \put(3.238,4.8){{\psfig{figure=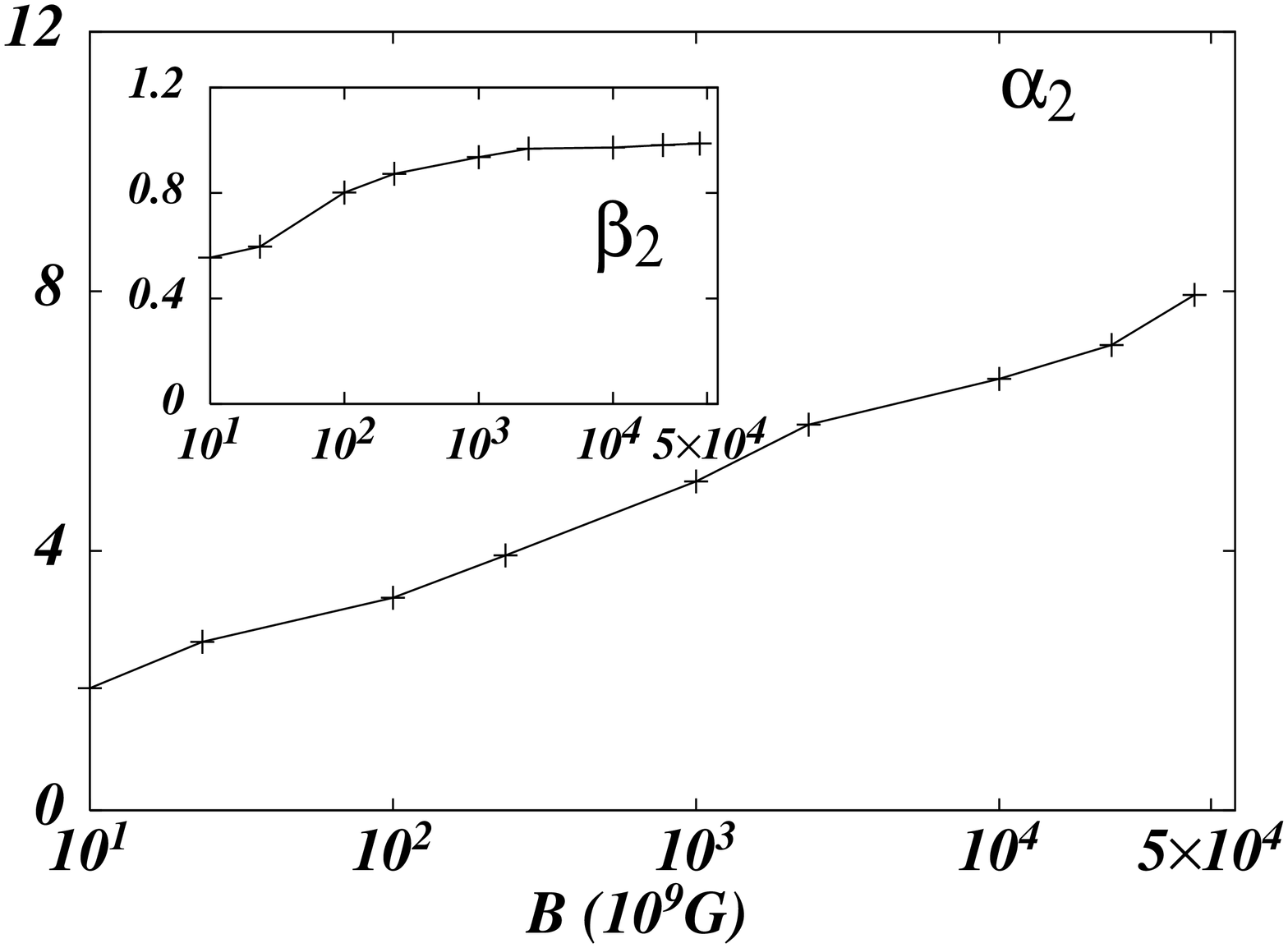,width=3.238in,angle=-90}}}
%%%%%%%%%%%%%%%%%%%%%%%%%%%%%%%%%%%%%%%%%%%%%%%%%%%%%%%%%%%%%%%%%%%%
   \end{picture} 
  \end{center} 
\caption{\label{fig:parms}  Variational parameters of the trial 
  function (\ref{psi123}) (see text) for the state $1\sigma_g$ at the
  equilibrium position as a function of the magnetic field strength
  $B$.  Parameters $\alpha_{1\ldots 10}$ are of dimension
  $[a.u.]^{-1}$ and the parameters $\beta_{1\ldots 6}$ are
  dimensionless. The parameter $A_1$ is placed equal to 1. In the
  figure where the parameters $\al_{8,9,10}$ are shown the right scale
  corresponds to the parameter $\alpha_{10}$.}
  \end{figure}

%%%%%%%%%%%%%%%%%%%%%%%%%%%
\addtocounter{figure}{-1} 
%%%%%%%%%%%%%%%%%%%%%%%%%%%

% PART B %%%%%%%%%%%%%%%%%%%%%%%%%%%%%%%%
\begin{figure}[htbp]
\centerline{\psfig{figure=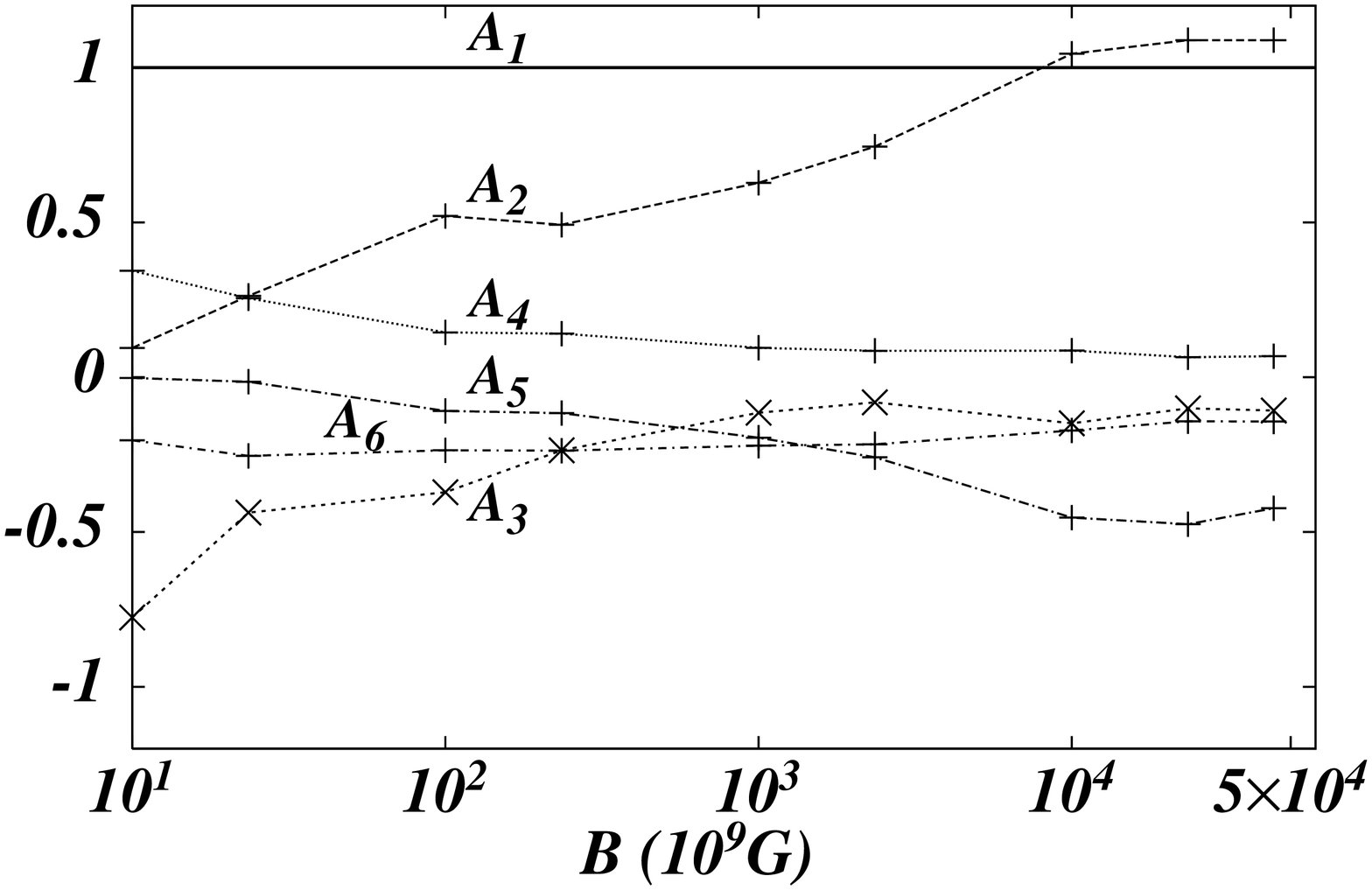,height=3in,angle=-90}}
  \caption{Continuation}
  \end{figure}

%%%%%%%%%%%%%%%%%   FIGURE:3 %%%%%%%%%%%%%%%%% 
%%%%%%%%%%%%%%  H3++ Transition Energies
\begin{figure}[tb]
 \unitlength=1in
 \begin{picture}(7.0,2.4)(0,0)
 \put(1.5,0){{\psfig{figure=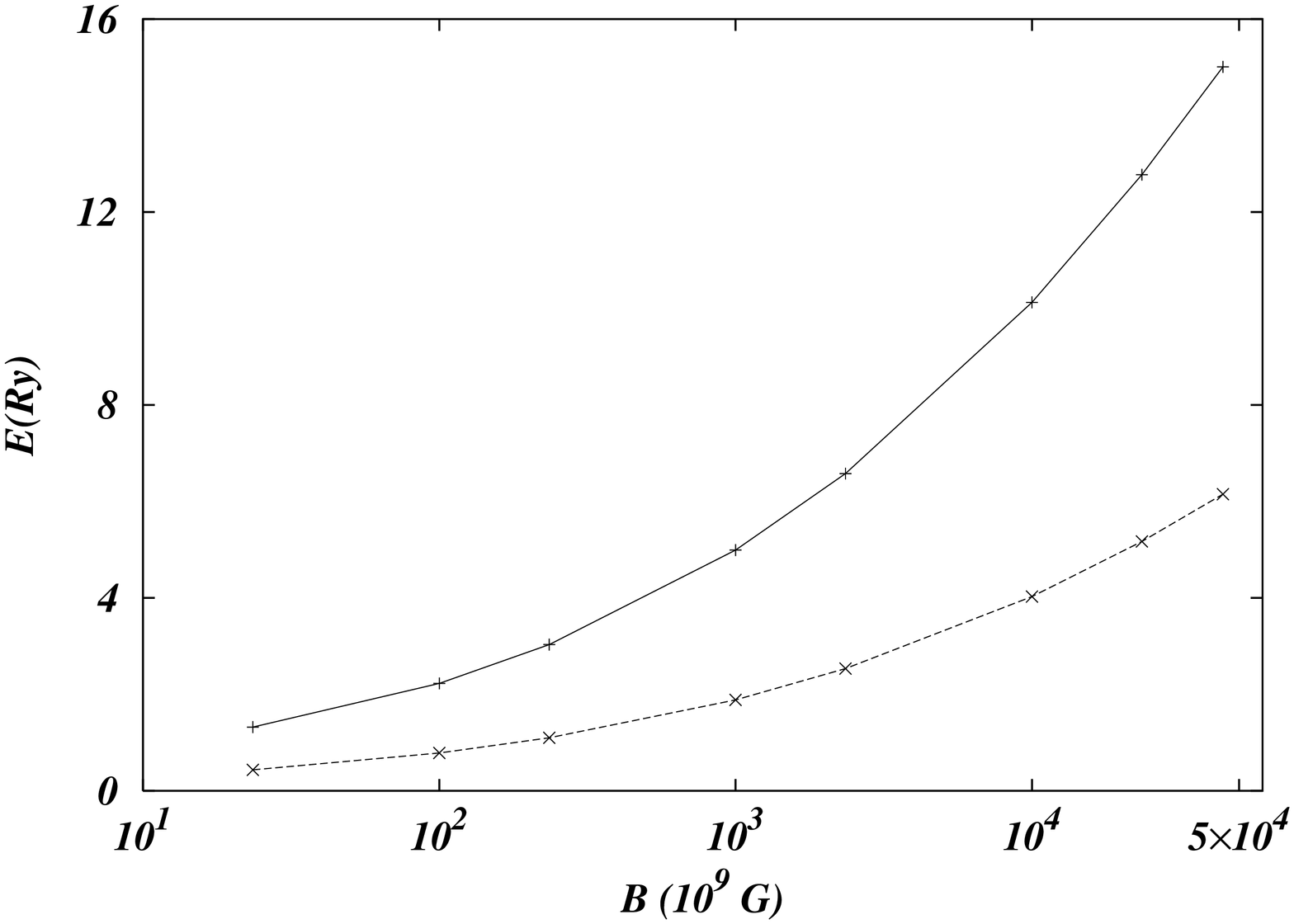,width=4in,angle=-90}}}
 \put(3.9,2){$1\si_g \to 1\pi_u$}
 \put(4.3,1.2){$1\pi_u \to 1\de_g$}
 \end{picture}
    \caption{\label{fig:trans} Transition energies between the lowest
      three states of the $H_3^{2+}$ ion as function of the magnetic
      field strength.}
\end{figure}
%%%%%%%%%%%%%% $E_{1 \pi_u} - E_{1 \sigma_g}$
%%%%%%%%%%%%%% $E_{1 \delta_g} - E_{1 \pi_u}$

%%%%%%%%%%%%%%%%%   FIGURE:4 %%%%%%%%%%%%%%%%% 
% PART A %%%%%%%%%%%%%%%%%%%%%%%%%%%%%%%%
\begin{figure}[h] 
  \unitlength=1in 
  \begin{center}
%\fbox{ 
  \begin{picture}(7.0,7.6)(0,0)
  \put(0.0,0){{\psfig{figure=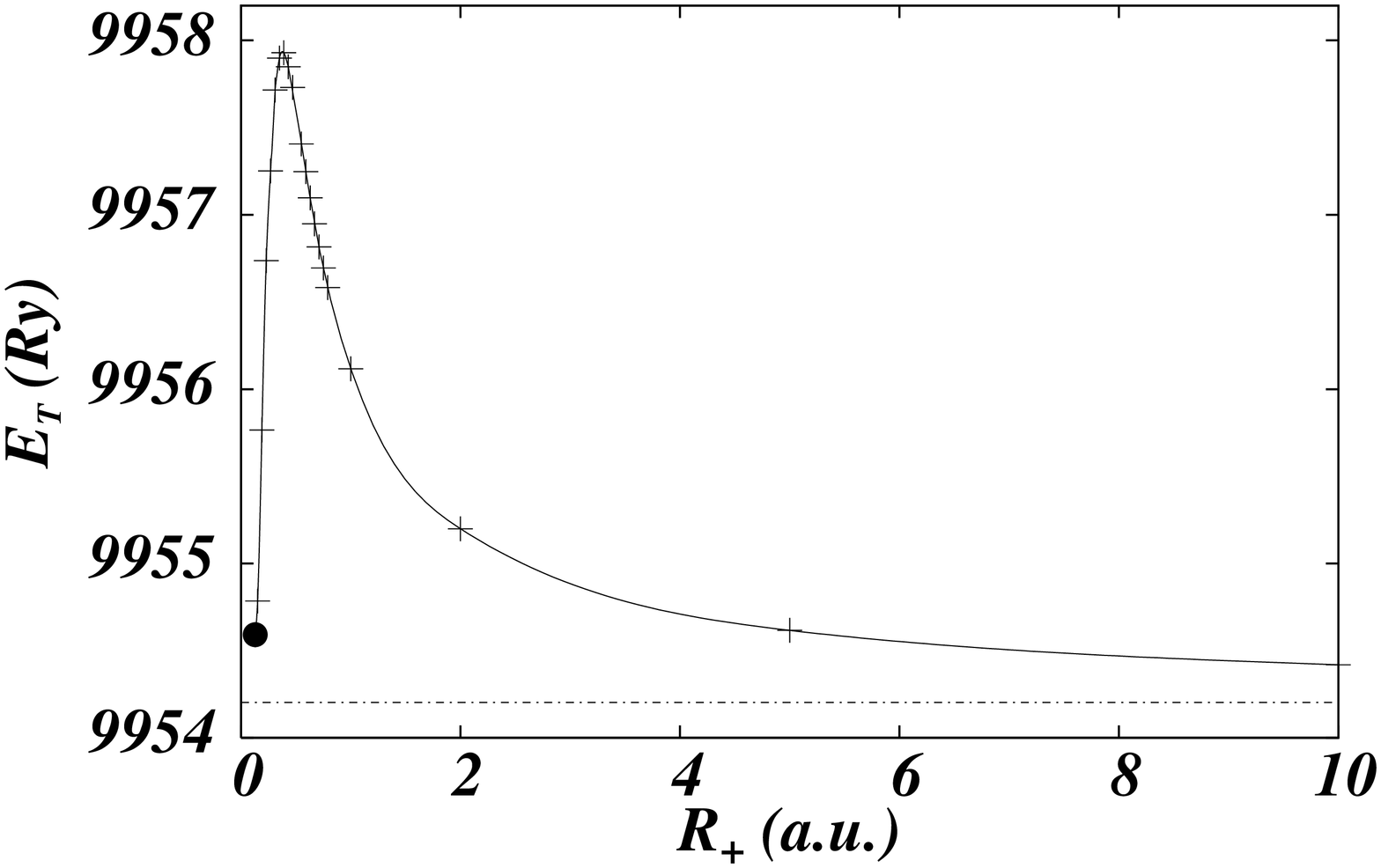,width=3.1in,angle=-90}}}
  \put(3.8,0){{\psfig{figure=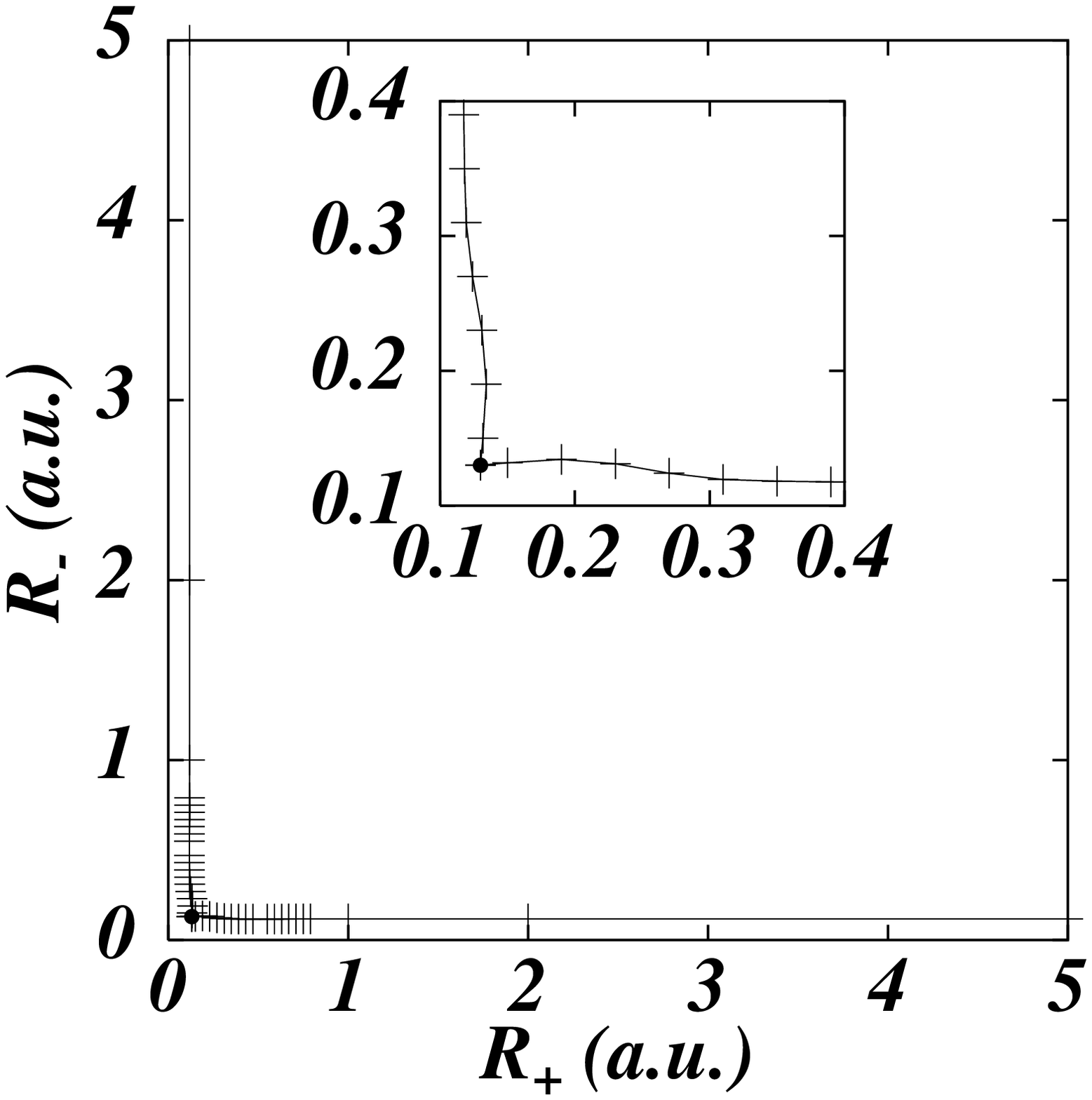,width=3.1in,angle=-90}}}
  \put(0.0,2){{\psfig{figure=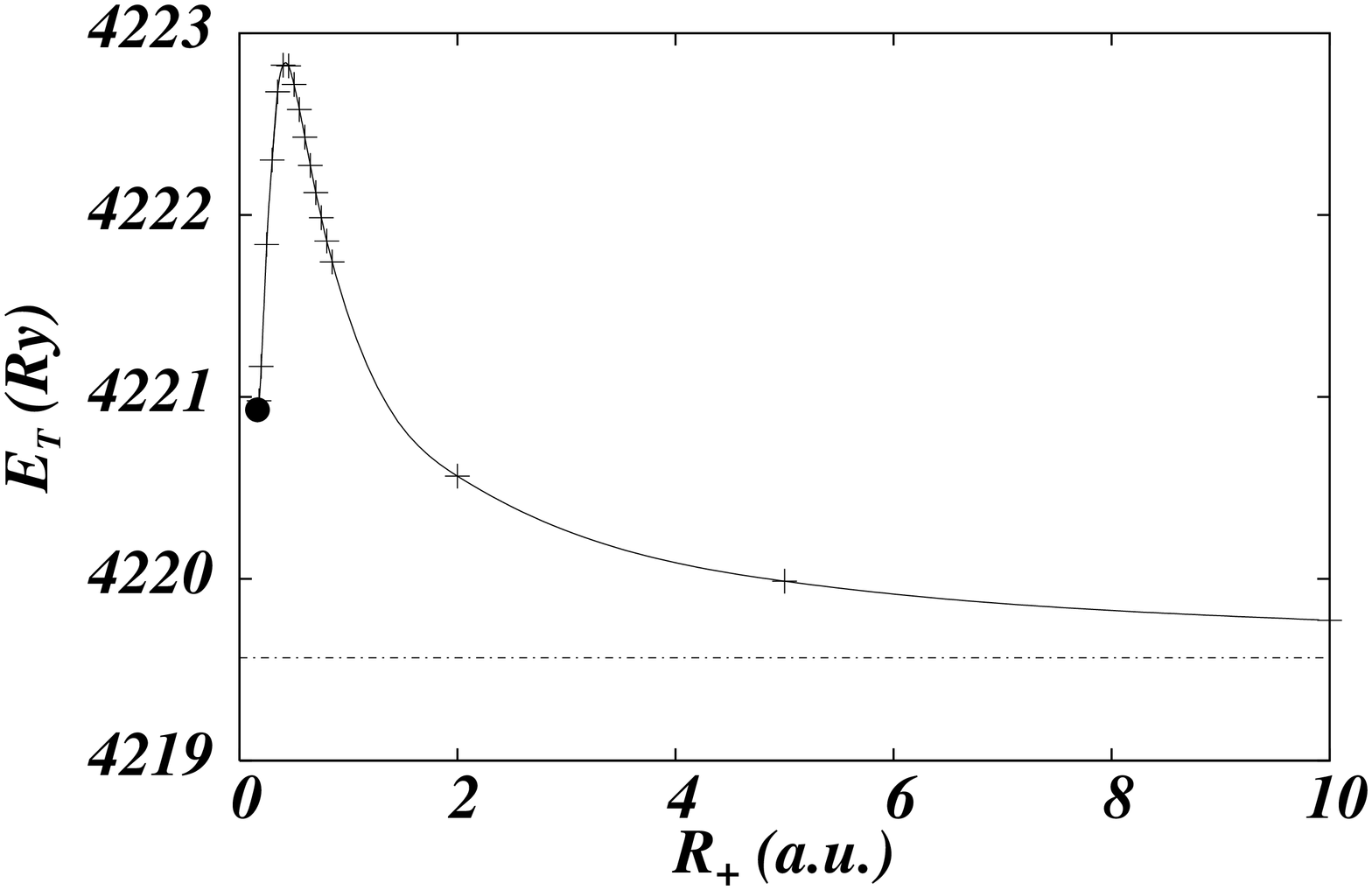,width=3.1in,angle=-90}}}
  \put(3.8,2){{\psfig{figure=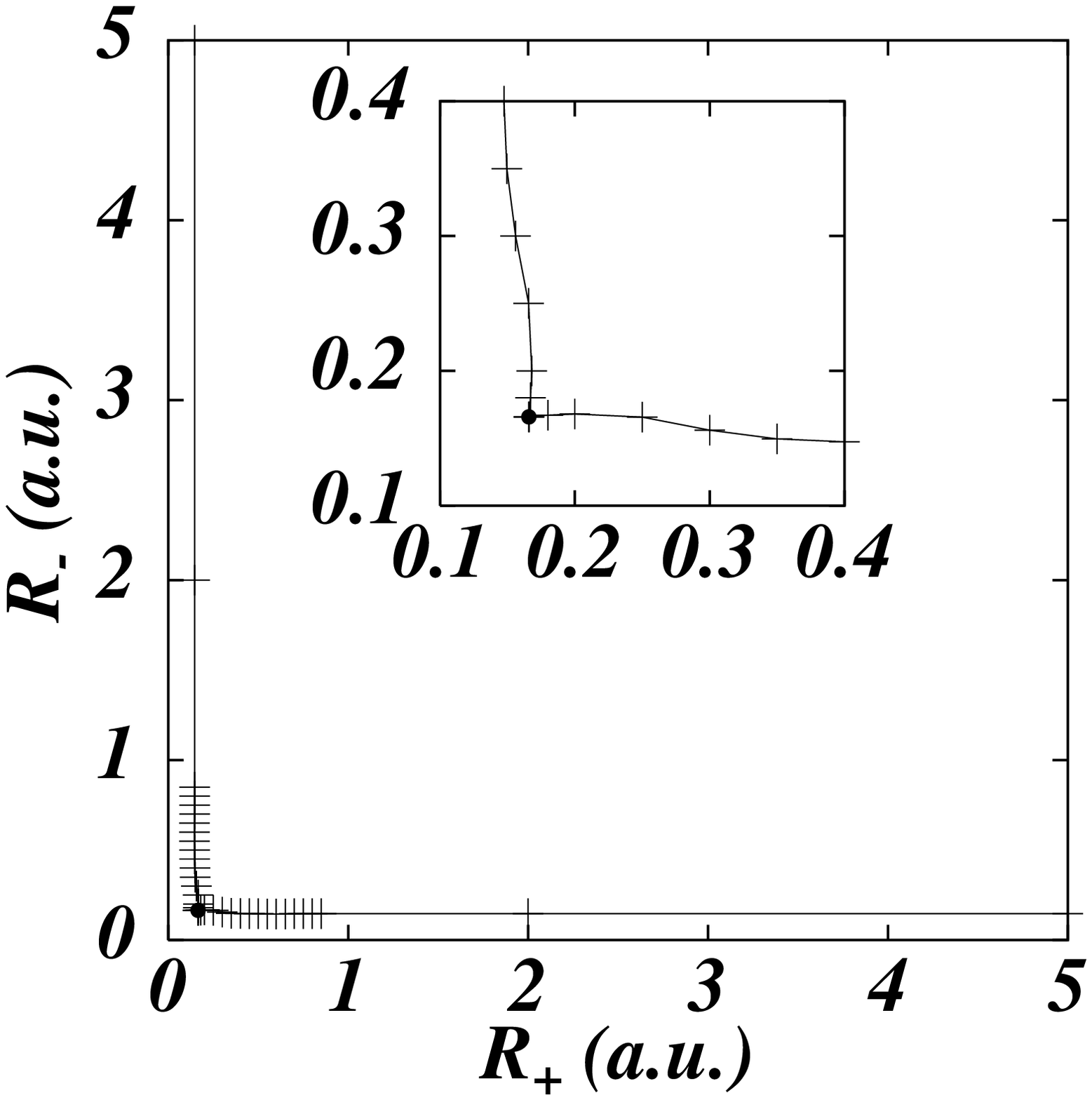,width=3.1in,angle=-90}}}
  \put(0,4){{\psfig{figure=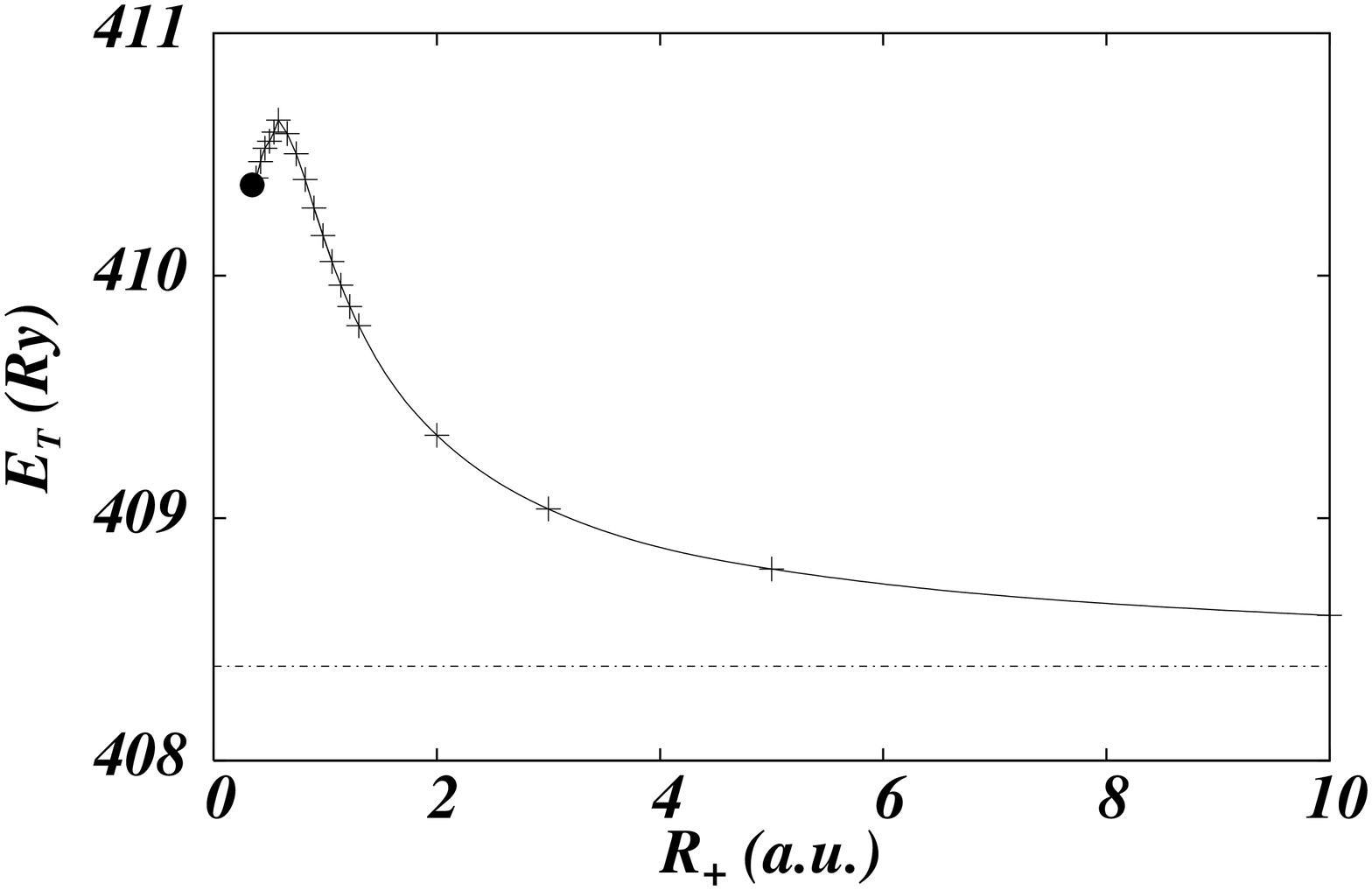,width=3.1in,angle=-90}}}
  \put(3.8,4){{\psfig{figure=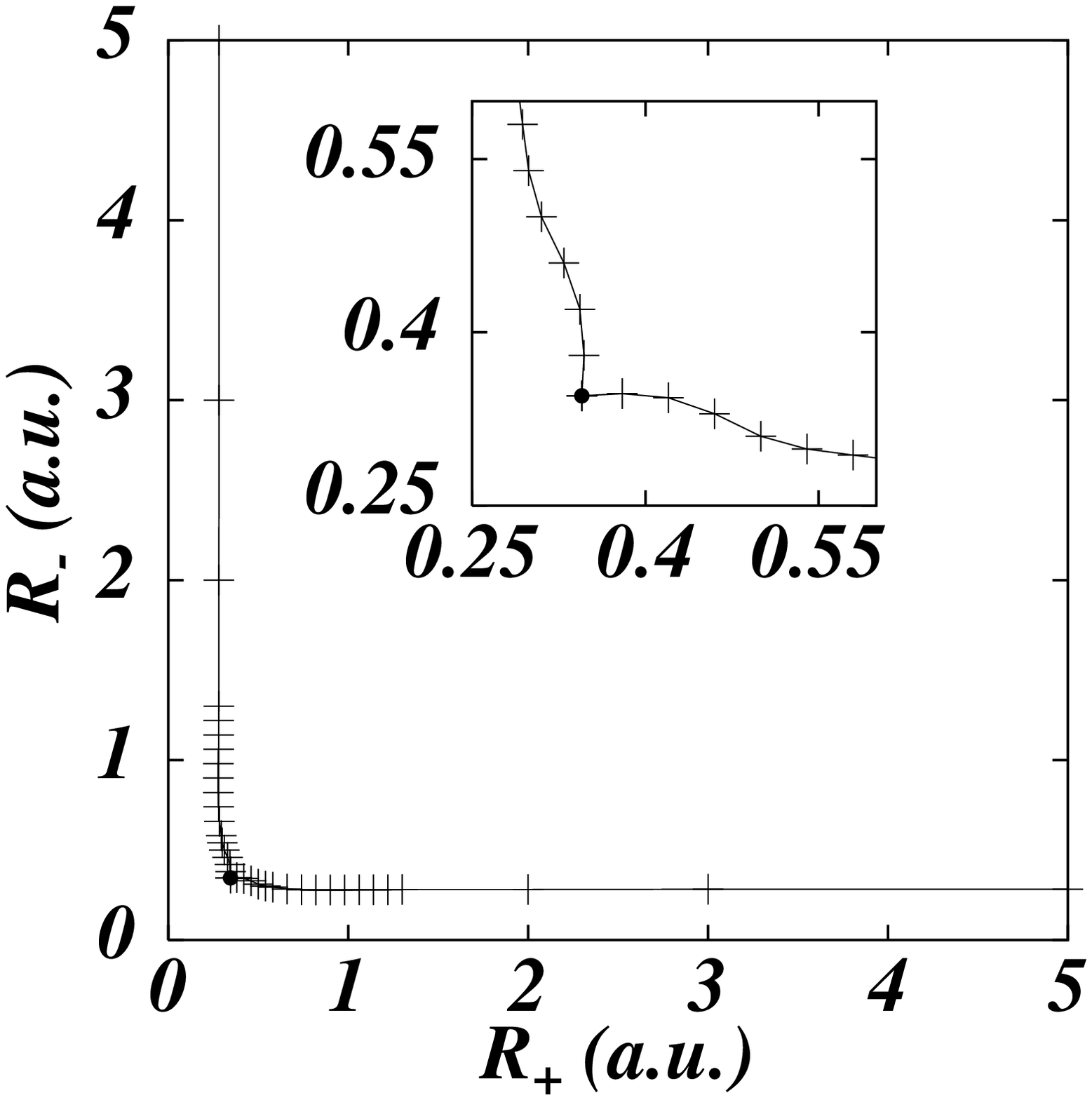,width=3.1in,angle=-90}}}
  \put(0,6){{\psfig{figure=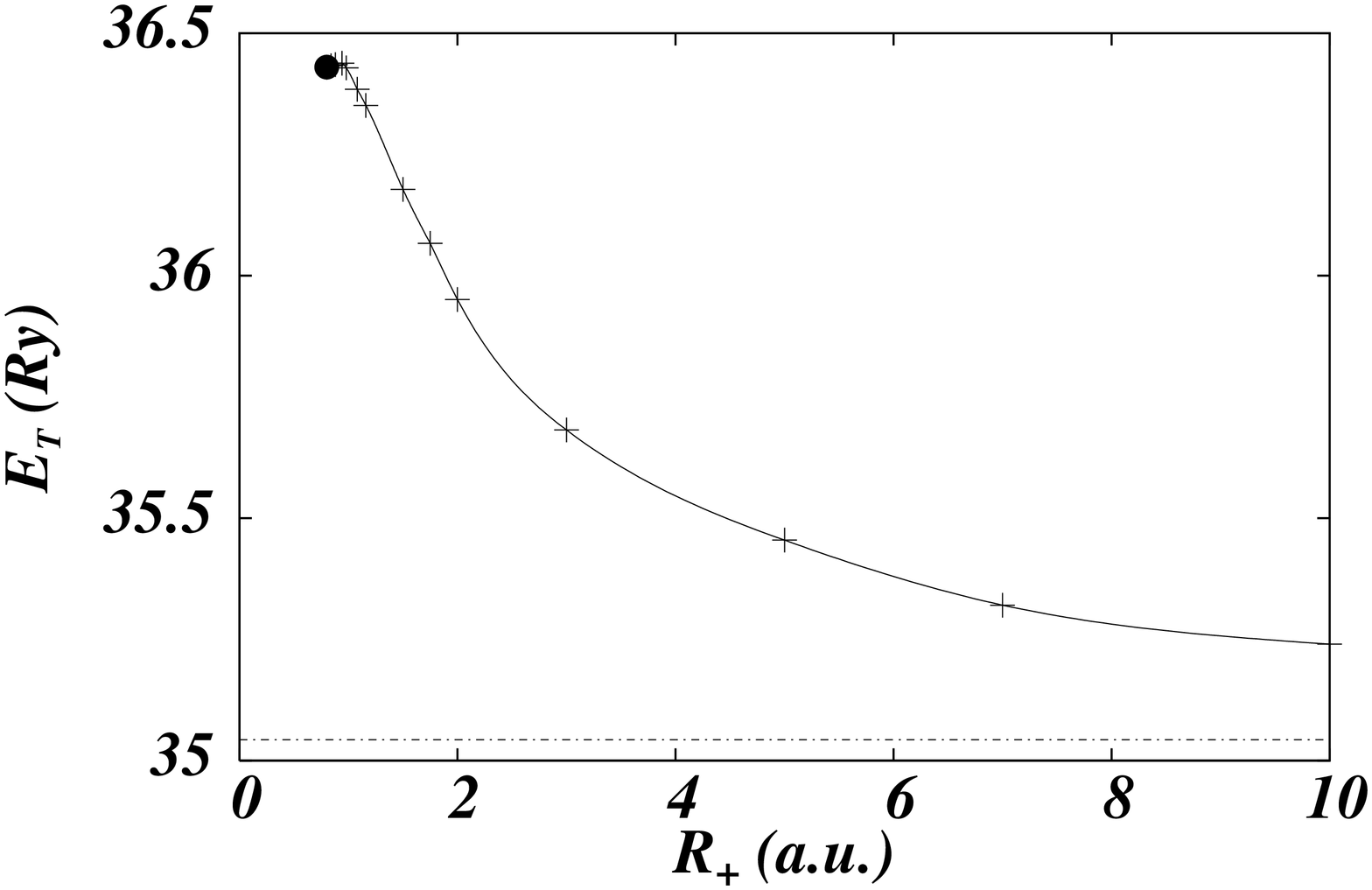,width=3.1in,angle=-90}}}
  \put(3.8,6){{\psfig{figure=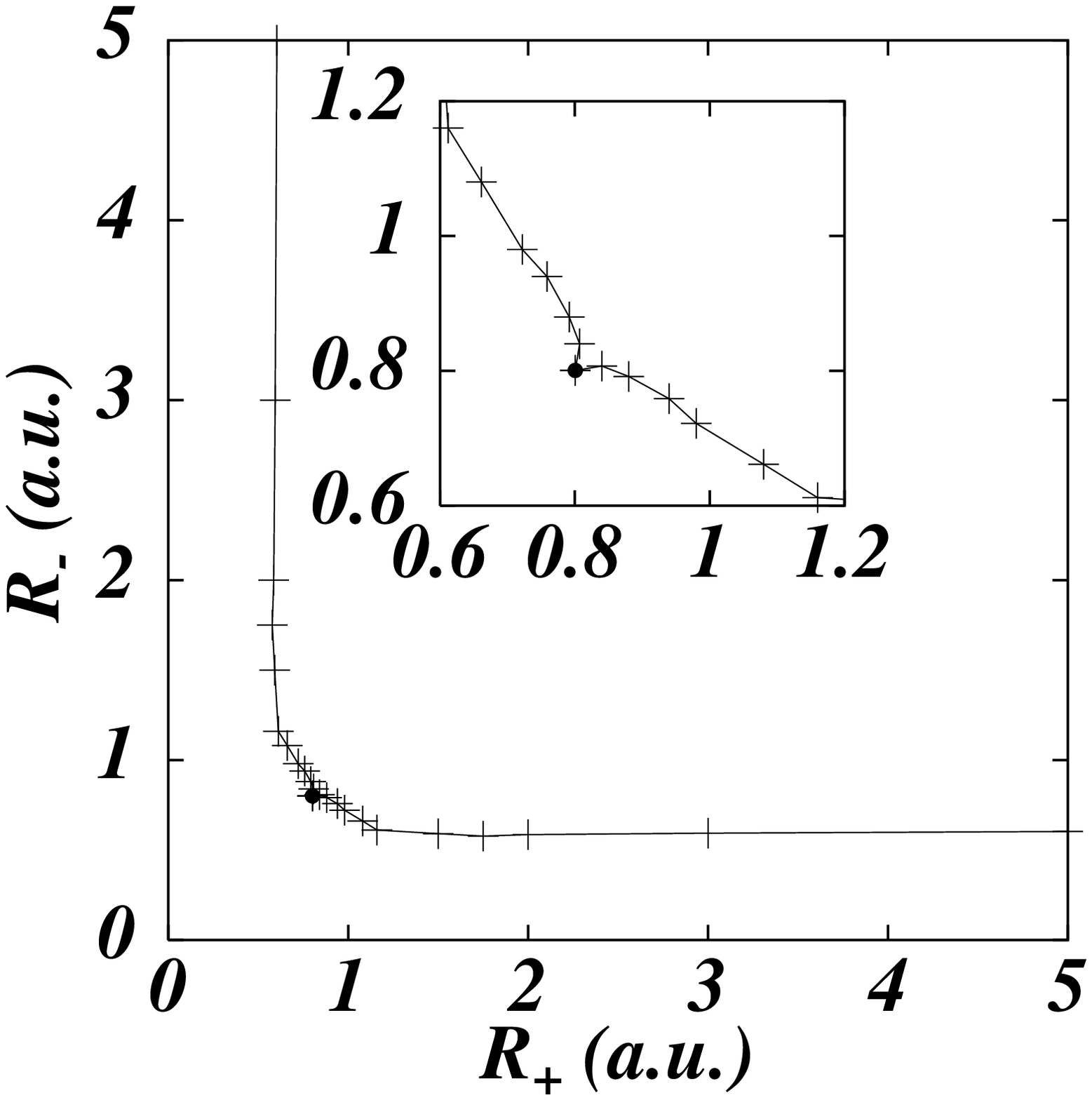,width=3.1in,angle=-90}}}
%%%%%%%% labels
   \put(3.3,7){(a)}
   \put(3.3,5){(b)}
   \put(3.3,3){(c)}
   \put(3.3,1){(d)}
%  \put(3.3,7){(e)}
   \end{picture}
%}
  \end{center}
\caption{\label{fig:valleys}Energy profiles  along the valleys of minimal total
  energy for the $H_3^{2+}$ ion as function of $R_+,R_-$, and the
  corresponding paths in the plane $(R_+,R_-)$ for different magnetic
  fields: (a) $B=10^{11}$\,G , (b) $B=10^{12}$\,G , (c) $B=10^{13}$\,G
  , (d) $B=2.35\times 10^{13}$\,G, and (e) $B=4.414\times 10^{13}$\,G.
  The position of the minimum is indicated by a bullet. The horizontal
  dashed line in the energy profile curve represents the energy of the
  $H_2^+$ ion.}
\end{figure}

%%%%%%%%%%%%%%%%%%%%%%%%%%%
\addtocounter{figure}{-1}
%%%%%%%%%%%%%%%%%%%%%%%%%%%

% PART B %%%%%%%%%%%%%%%%%%%%%%%%%%%%%%%%
\begin{figure}[h]
  \unitlength=1in
  \begin{center}
%\fbox{
  \begin{picture}(7.0,2.4)(0,0)
  \put(0,0){{\psfig{figure=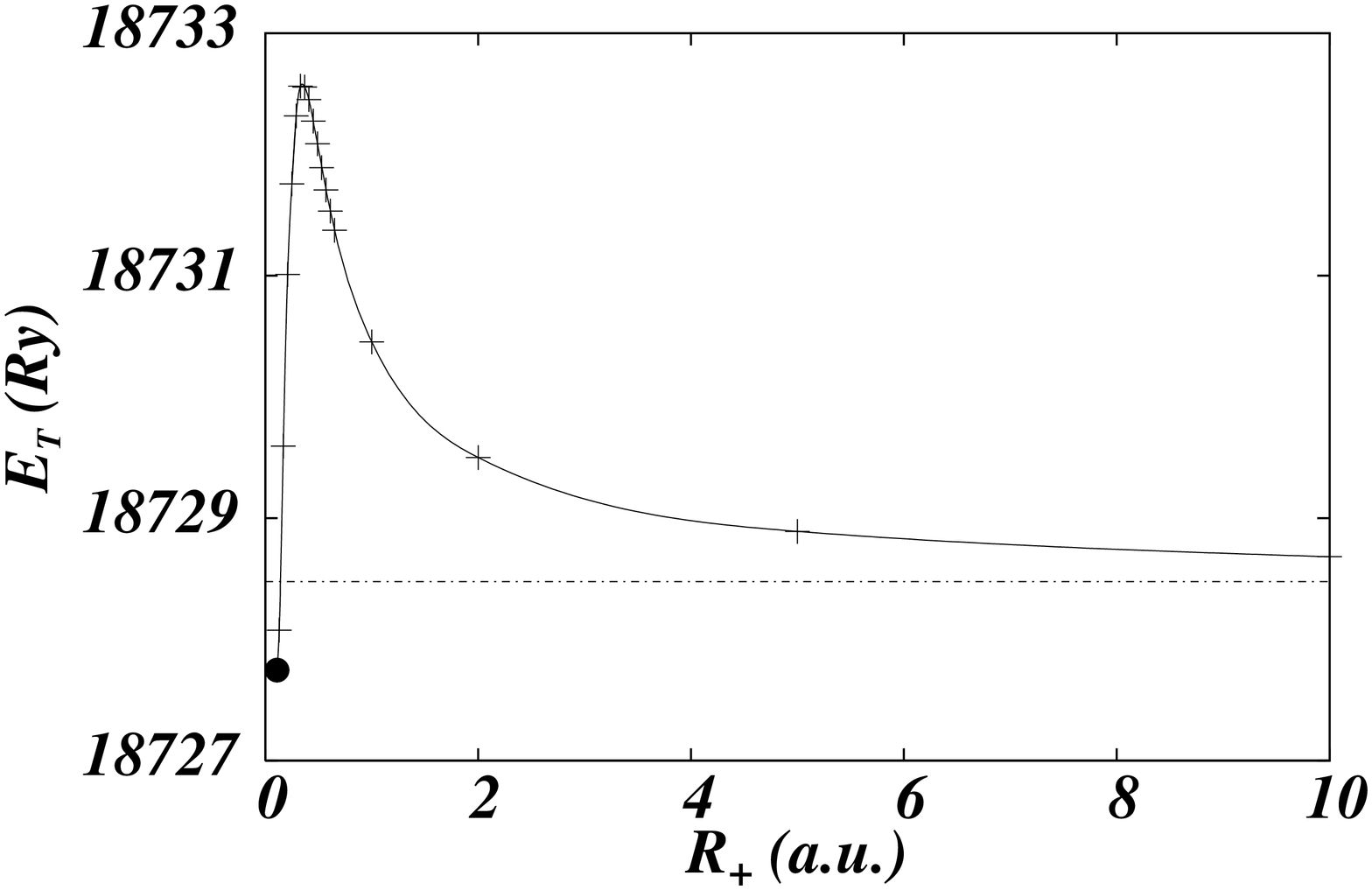,width=3.1in,angle=-90}}}
  \put(3.8,0){{\psfig{figure=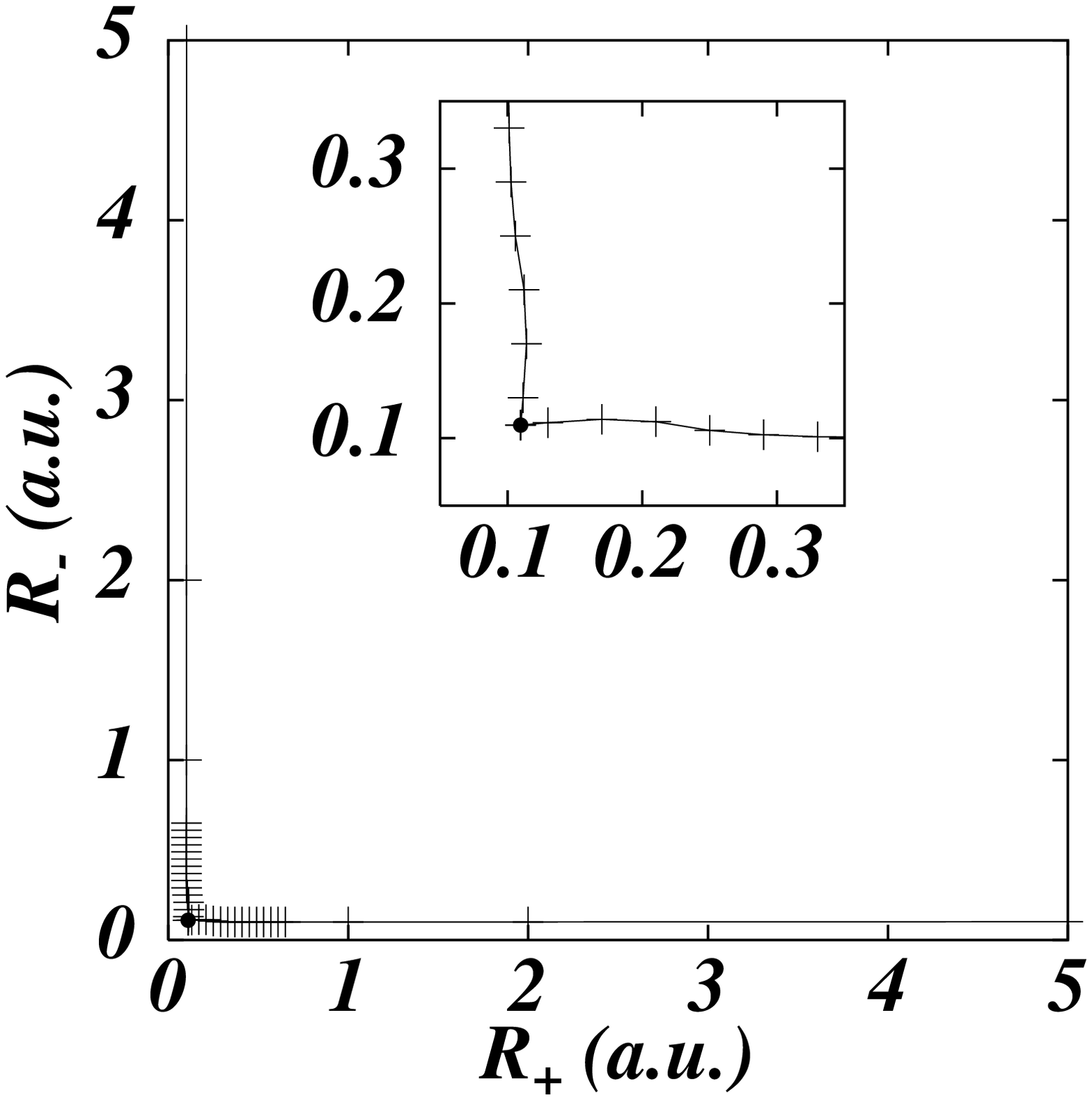,width=3.1in,angle=-90}}}
  \put(3.3,1){(e)}
   \end{picture}
%}
  \end{center}
\caption{Continuation}
\end{figure}

%%%%%%%%%%%%%%%%%   FIGURE:5 %%%%%%%%%%%%%%%%%
\begin{figure}[h]
  \unitlength=1in  
  \begin{center}
%\fbox{              
  \begin{picture}(7.0,7.6)(0,0)
  \put(0.0,0){{\psfig{figure=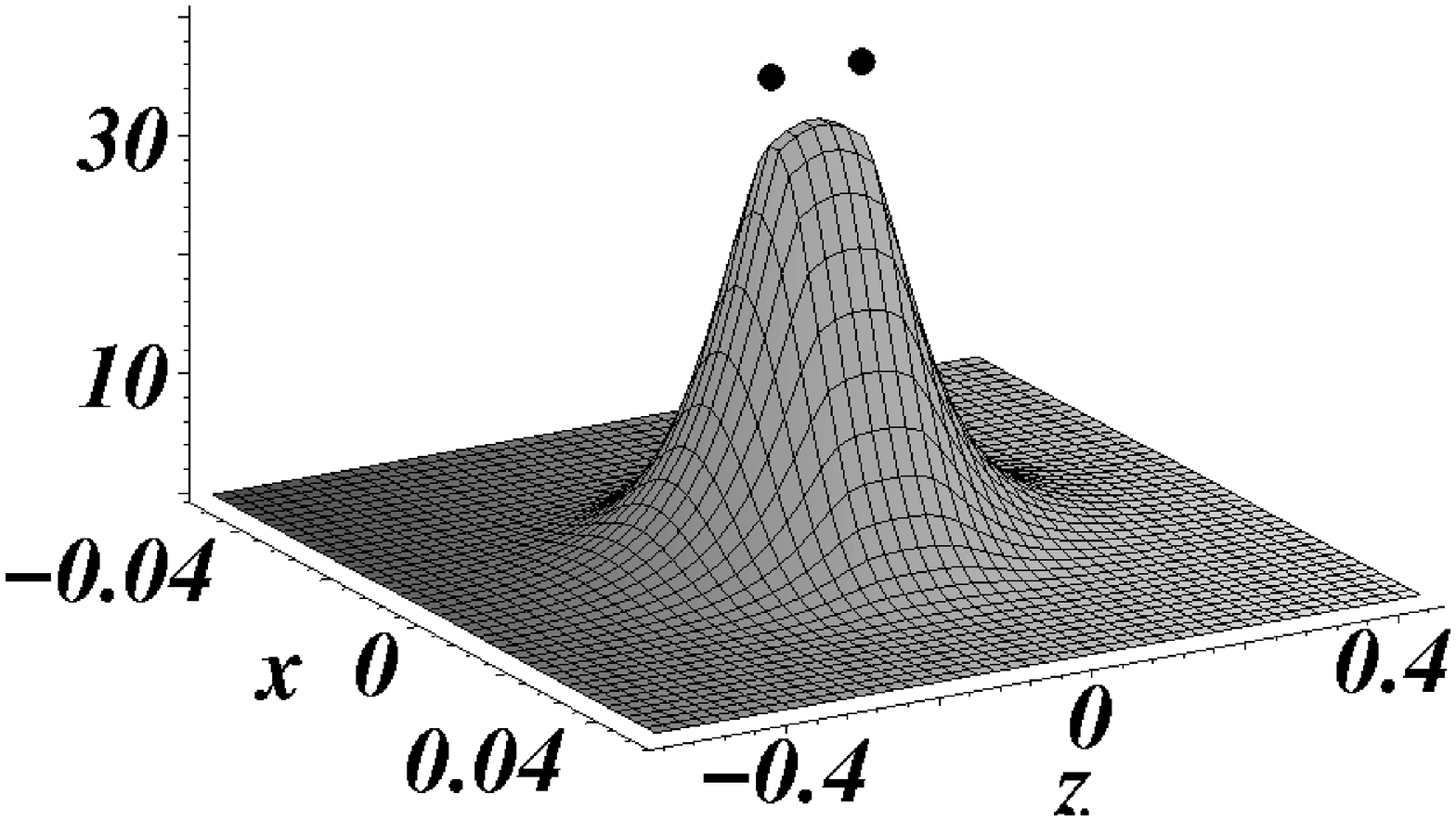,width=2.8in,angle=-90}}}
  \put(3.6,0){{\psfig{figure=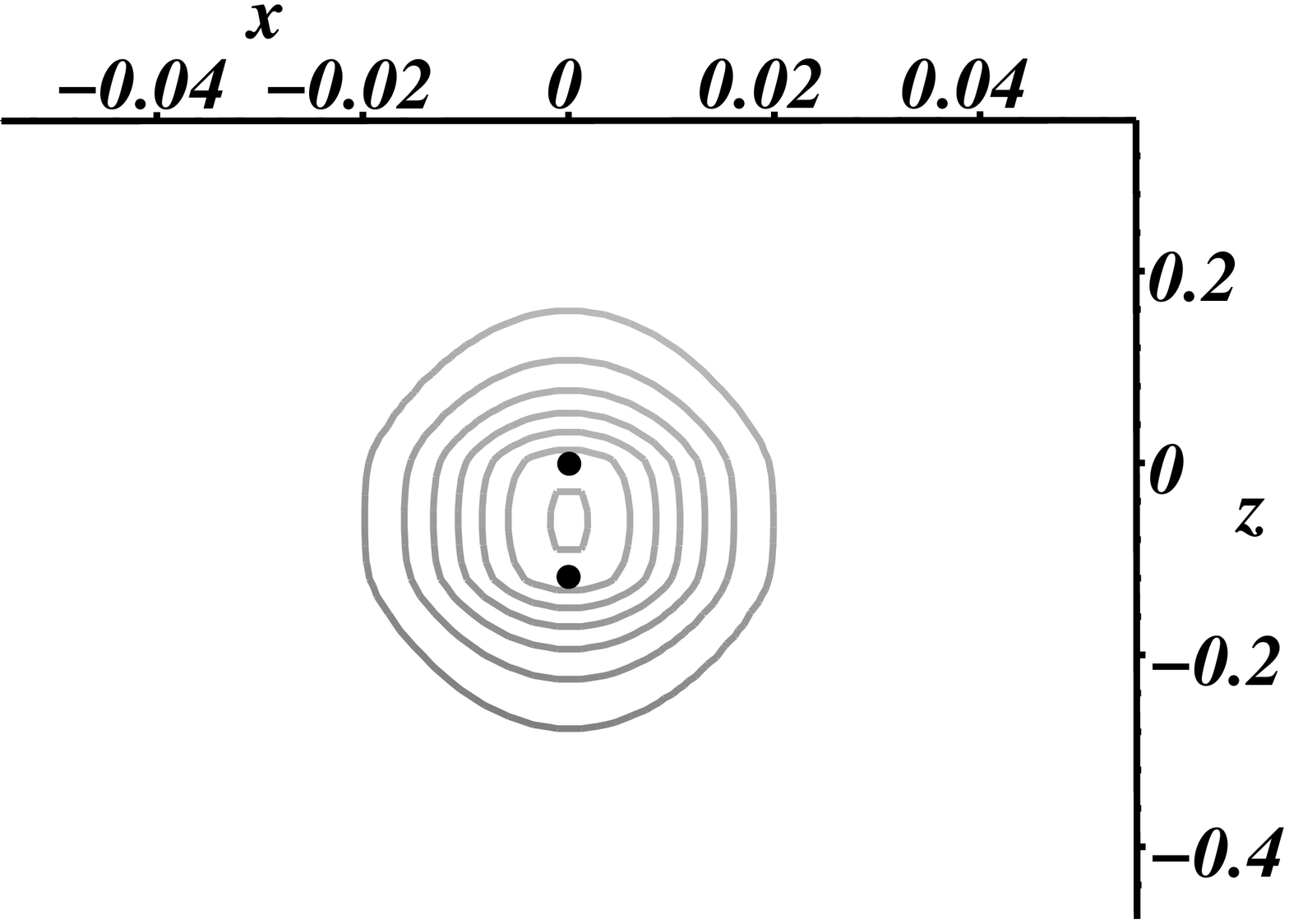,width=2.8in,angle=-90}}}
  \put(0.0,2){{\psfig{figure=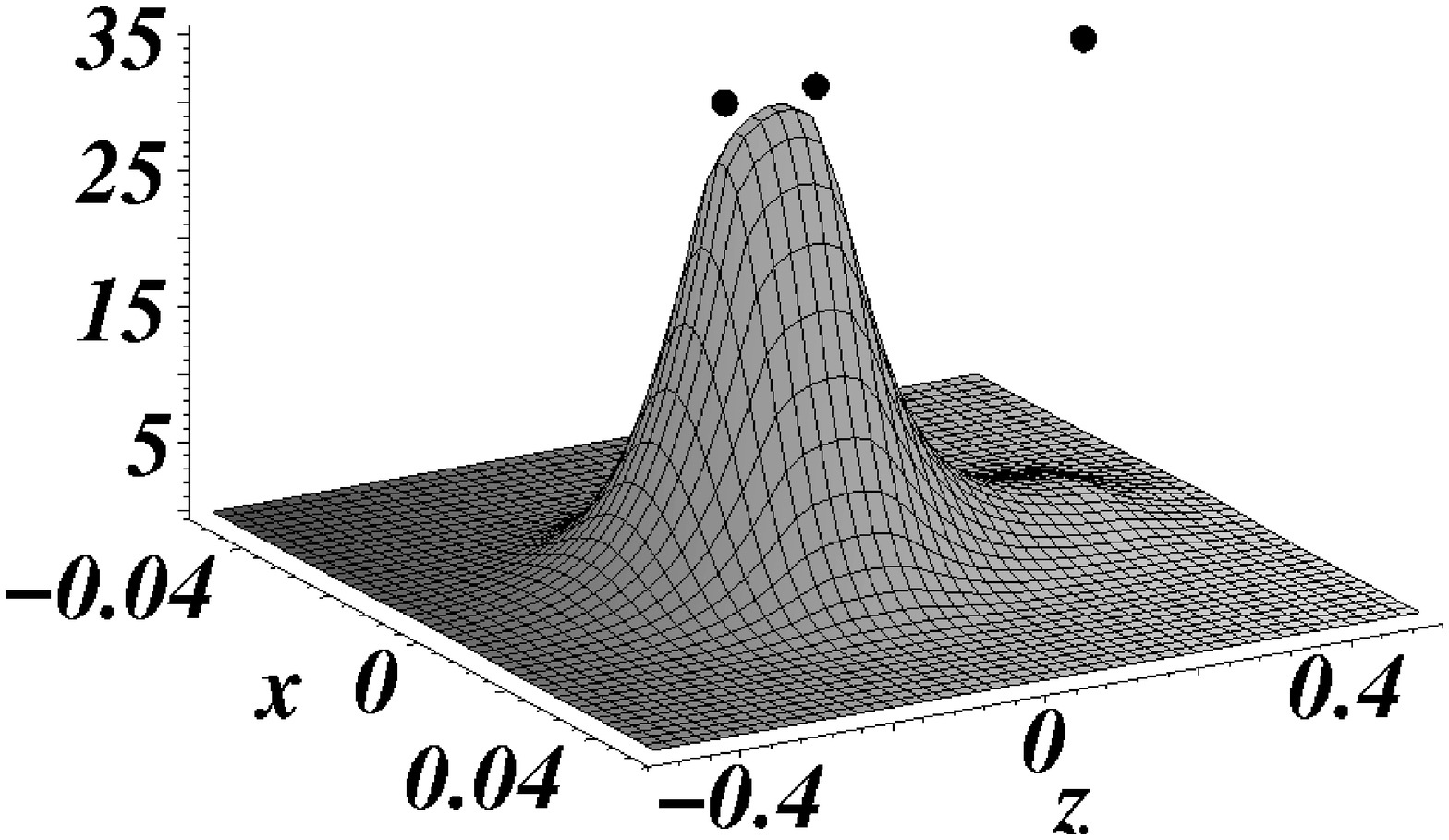,width=2.8in,angle=-90}}}
  \put(3.6,2){{\psfig{figure=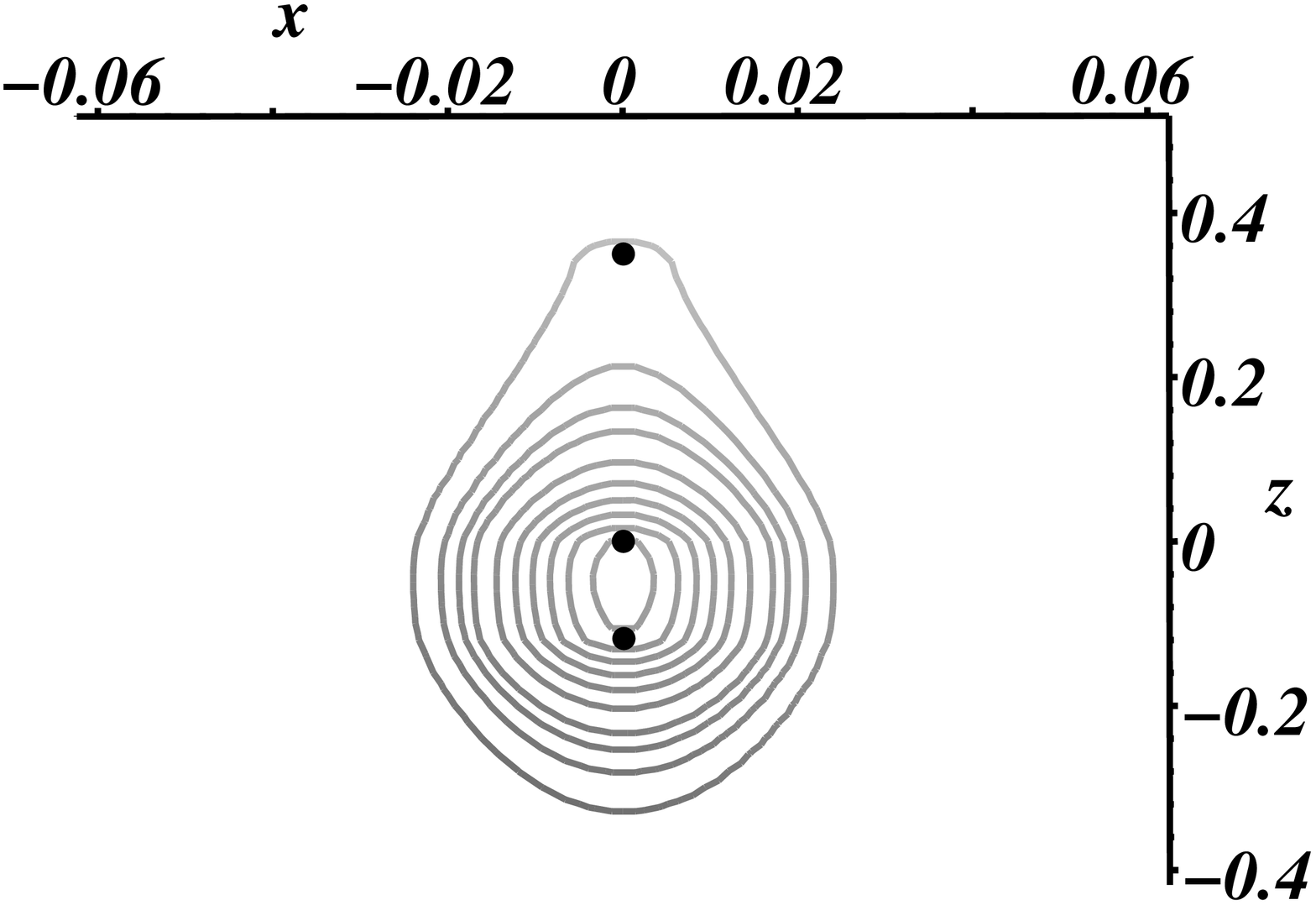,width=2.8in,angle=-90}}}
  \put(0,4){{\psfig{figure=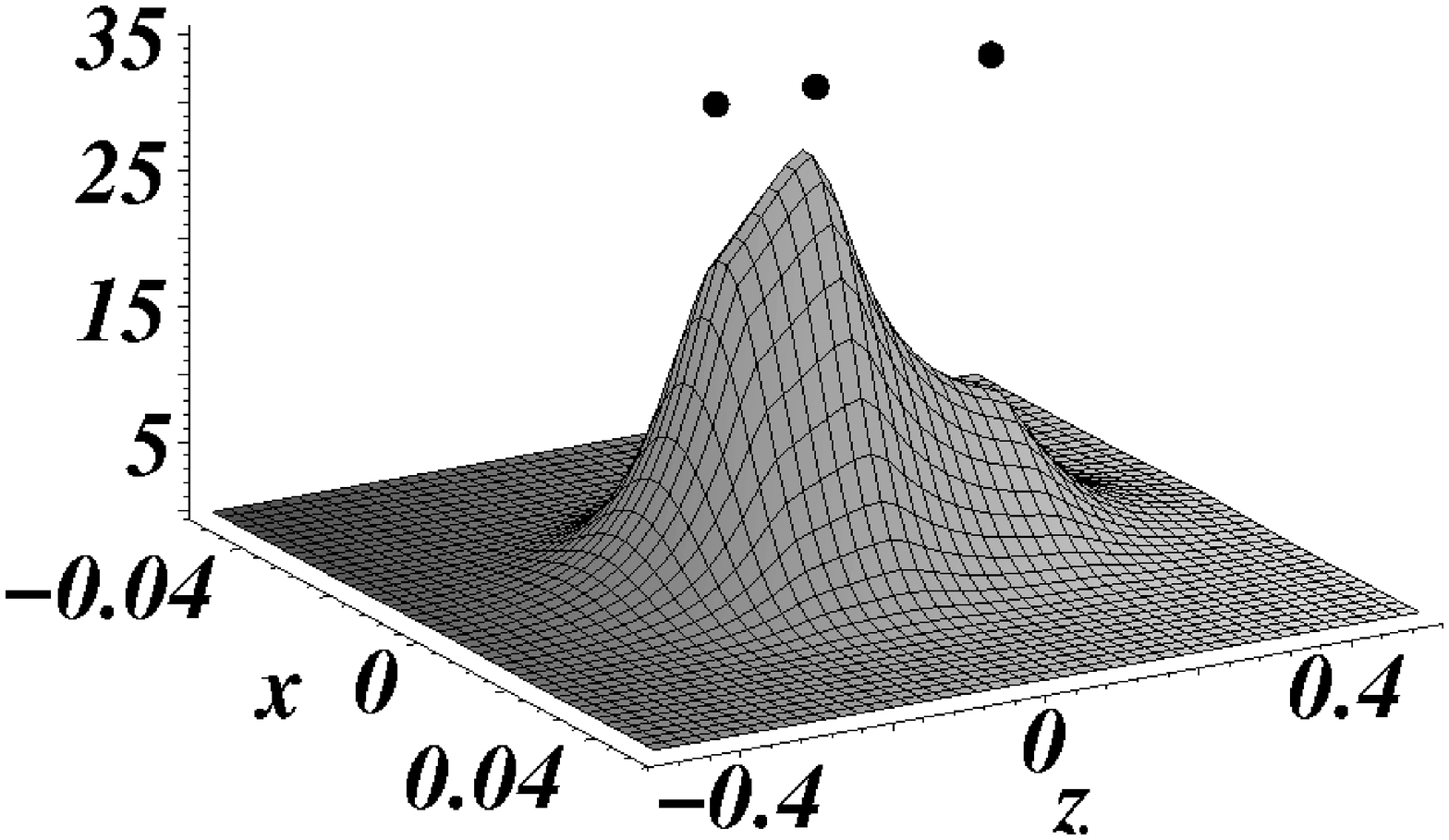,width=2.8in,angle=-90}}}
  \put(3.6,4){{\psfig{figure=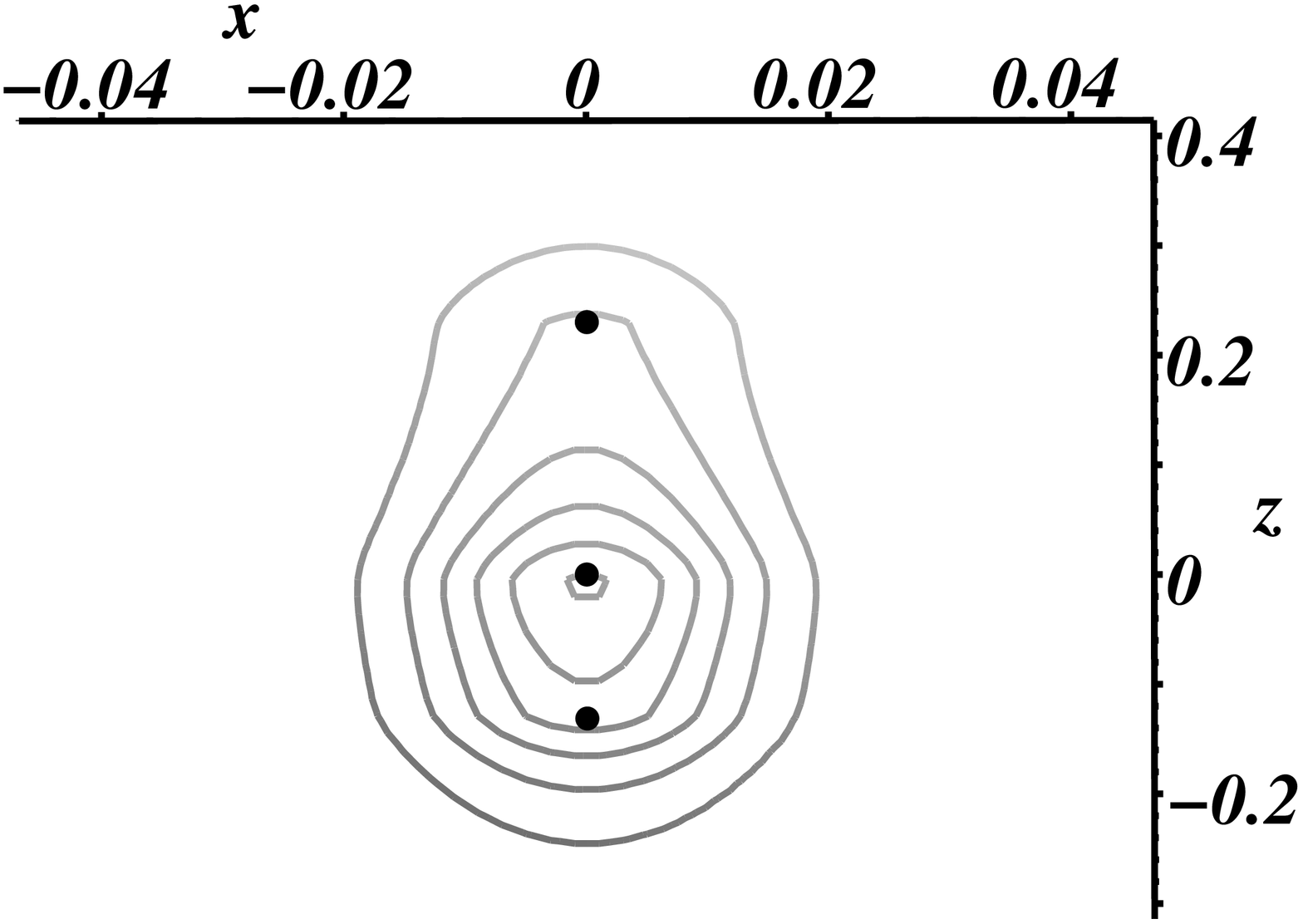,width=2.8in,angle=-90}}}
  \put(0,6){{\psfig{figure=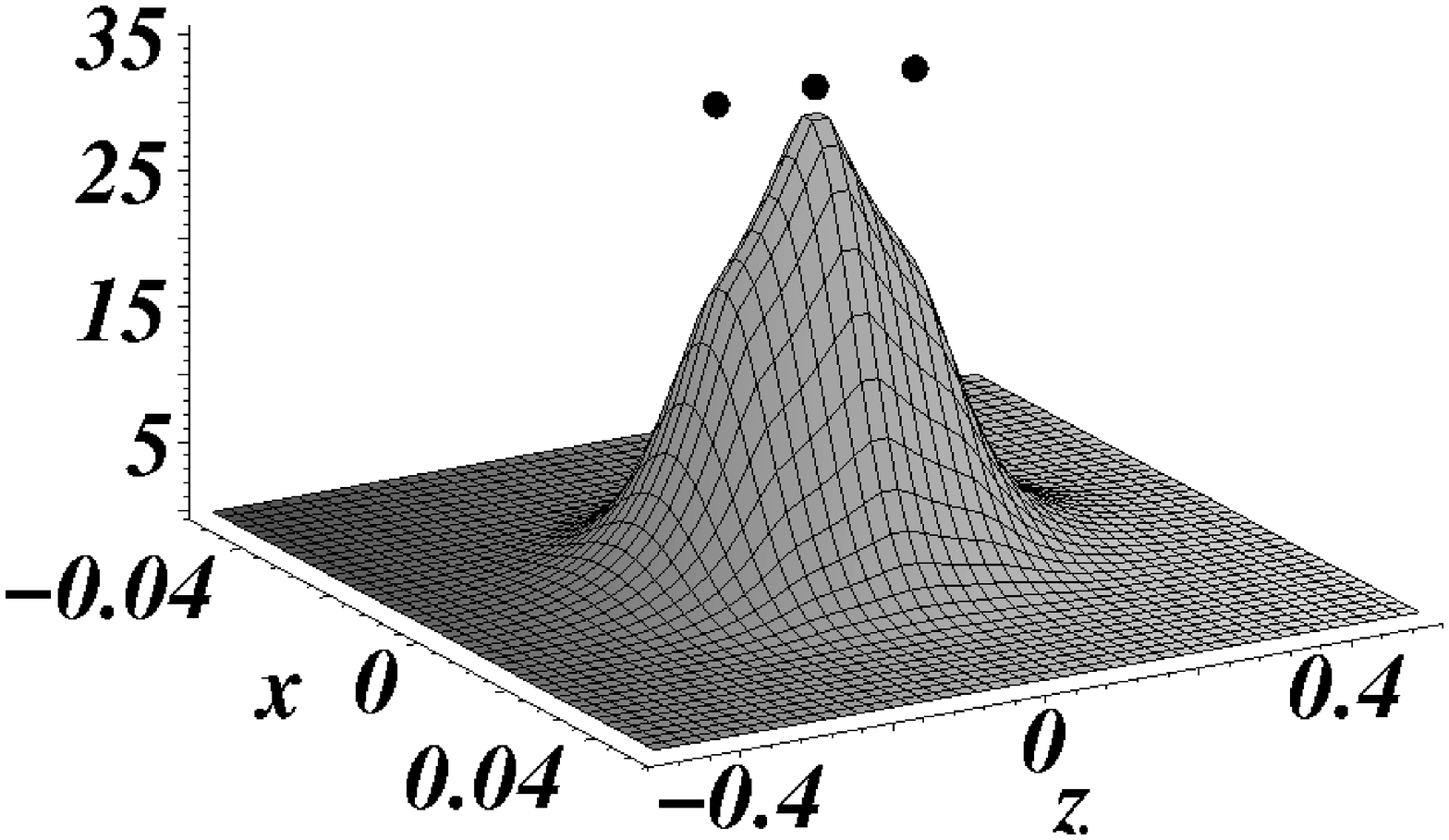,width=2.8in,angle=-90}}}
  \put(3.6,6){{\psfig{figure=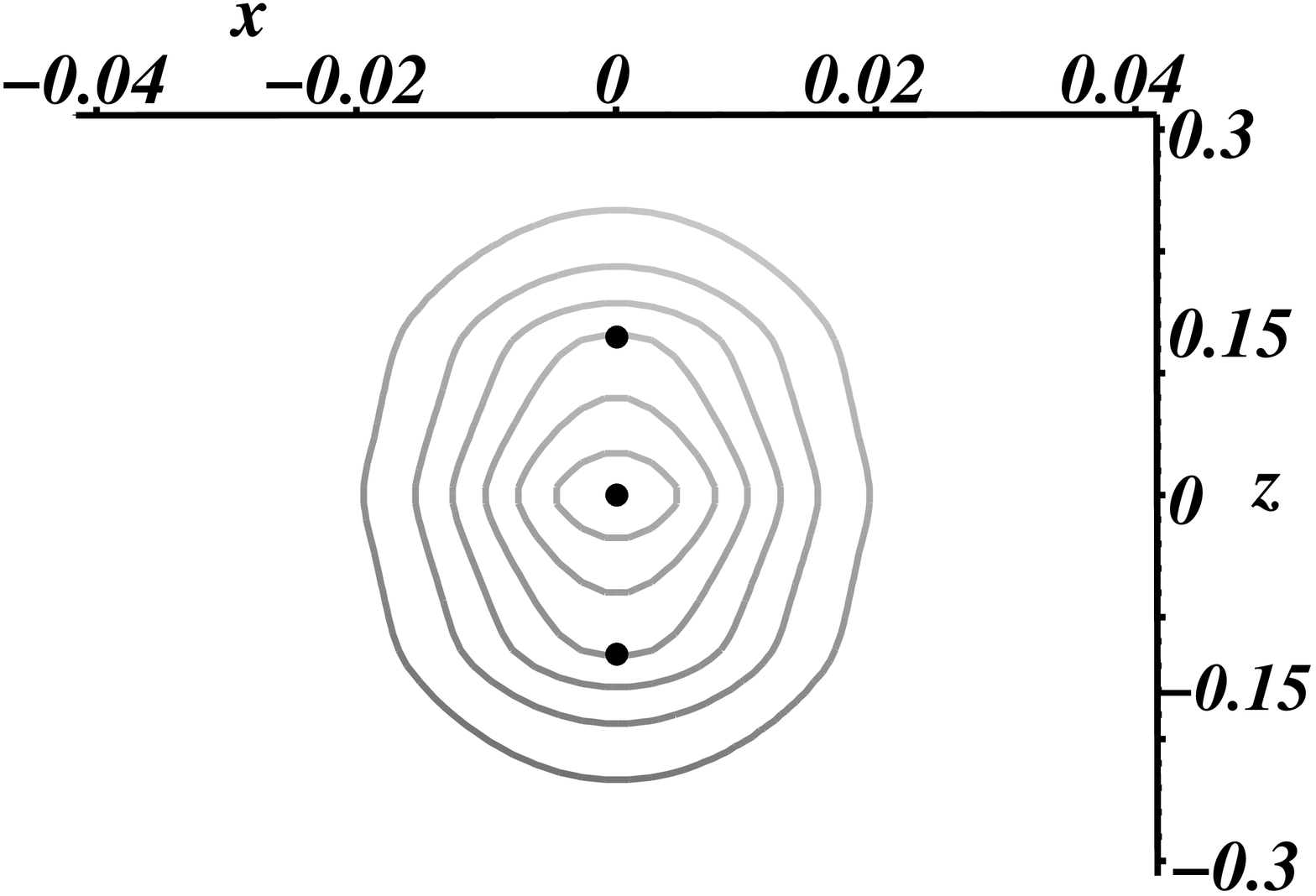,width=2.8in,angle=-90}}}
%%%%%%%%%%%%%%% 
  \put(0.39,7.3){$|\Psi|^2$}
  \put(0.39,5.3){$|\Psi|^2$}
  \put(0.39,3.3){$|\Psi|^2$} 
  \put(0.39,1.3){$|\Psi|^2$} 
%%%%%%%%%%%%%%%   
  \put(3,7.3){(a)} 
  \put(3,5.3){(b)} 
  \put(3,3.3){(c)}   
  \put(3,1.3){(d)}
   \end{picture}
%}
  \end{center}
\caption{\label{fig:edist1}Normalized electronic density distributions
  ${\Psi^2(x,y=0,z)}/{\int \Psi^2(x,y,z) d{\vec r}}$ and their
  contours for the ground state $1\si_g$ of the $H_3^{2+}$ ion along
  one of the valleys in a magnetic field $B=10000$\,a.u.: (a)
  $R_+=R_{eq}=0.13$\,a.u., (b) $R_+ = 0.23$\,a.u., (c) $R_+ =
  0.35$\,a.u. (near maximum, $R_+^{max}\simeq 0.36$\, a.u.), (d) $R_+
  = 5.0$\,a.u.  (third proton lies outside of the figure). The
  positions of the protons are marked by bullets. Vertical axis in the
  figures on the left is scaled to $1:1000$.}
\end{figure}    
     
%%%%%%%%%%%%%%%%%   FIGURE:6 %%%%%%%%%%%%%%%%%
\begin{figure}[h]
  \unitlength=1in
  \begin{center}  
%\fbox{
  \begin{picture}(7.0,5.2)(0,0)
%  \put(0.0,0)
%  \put(3.6,0)
  \put(0,0){{\psfig{figure=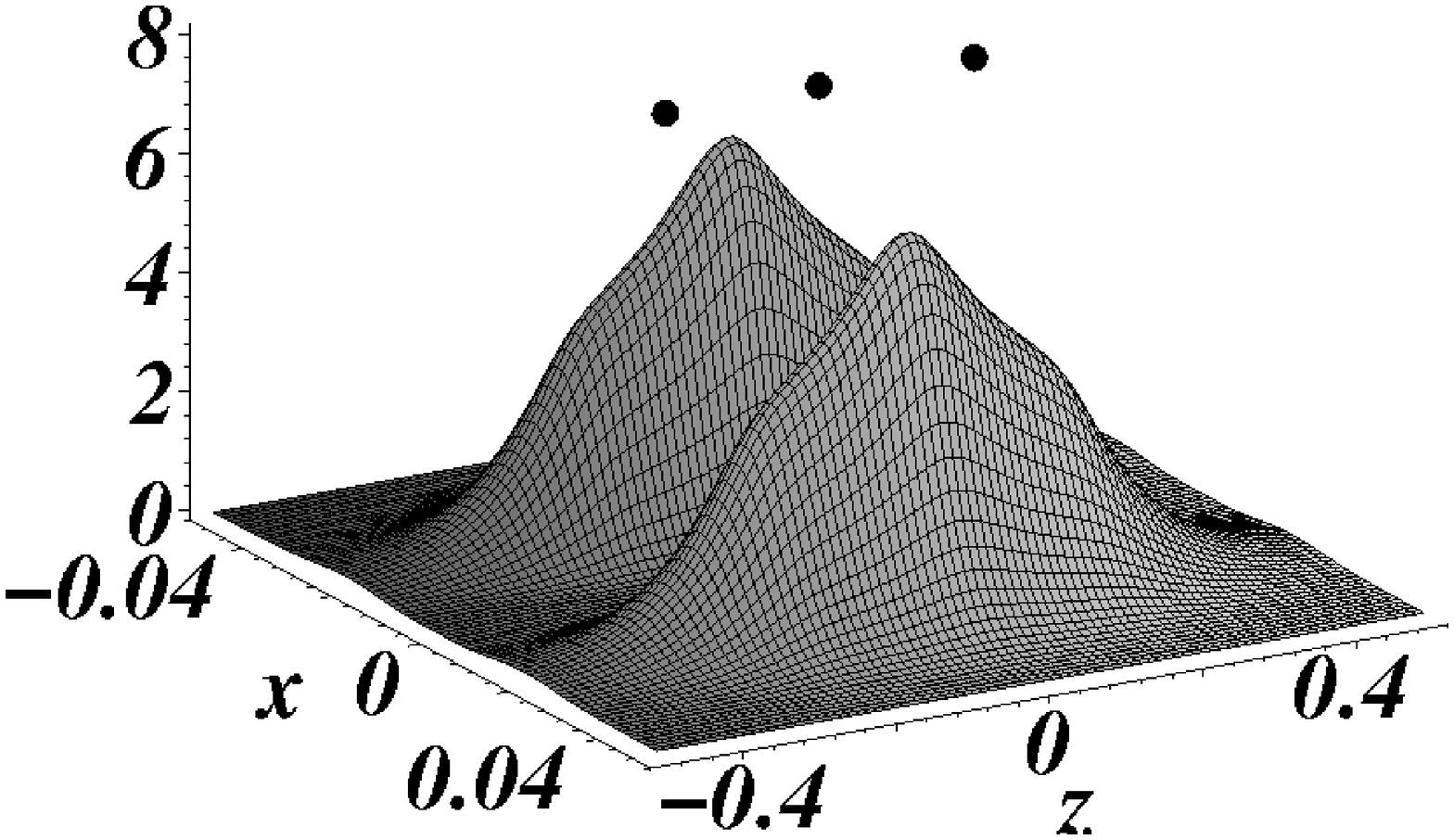,width=3in,angle=-90}}}
  \put(3.6,0){{\psfig{figure=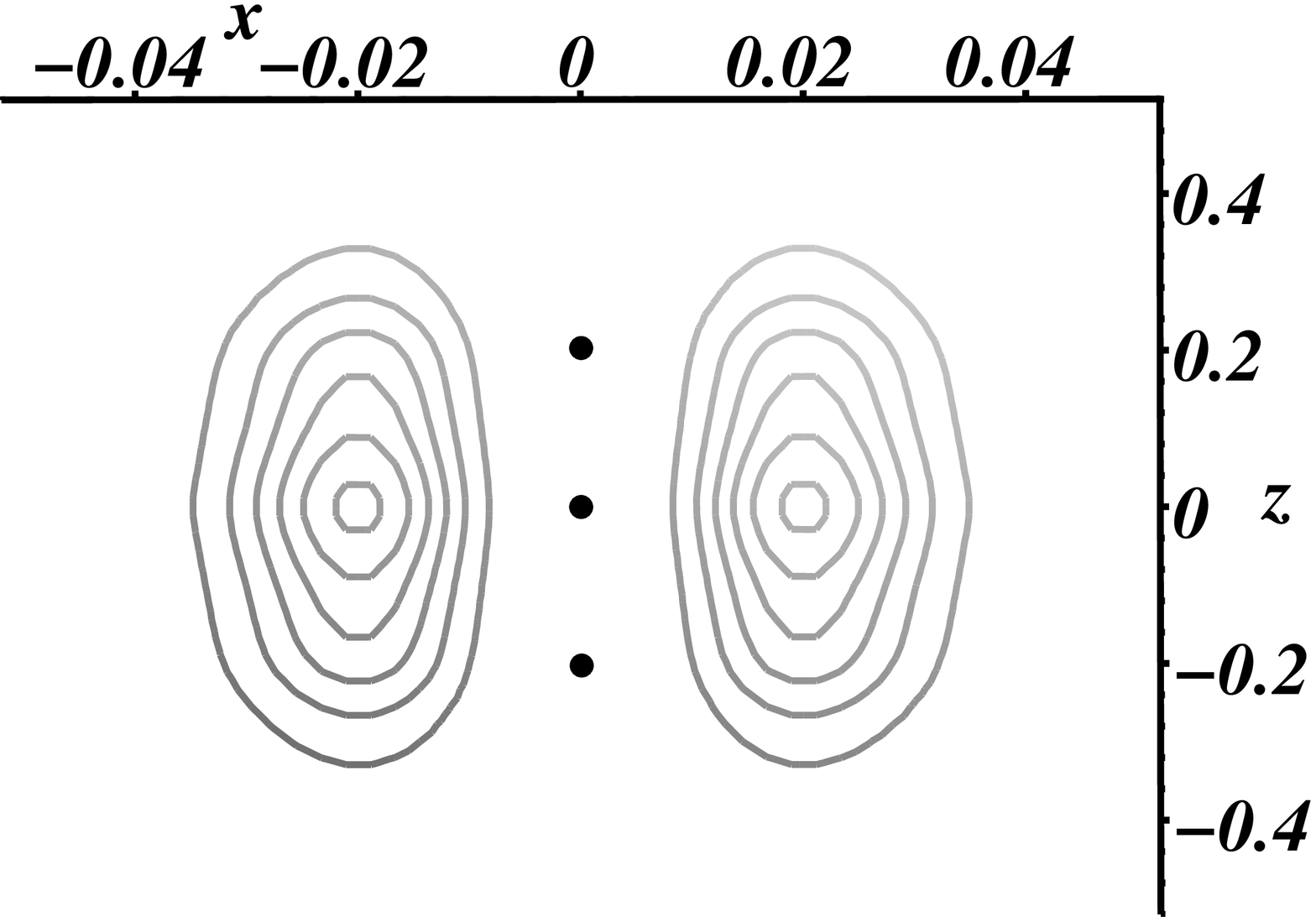,width=3in,angle=-90}}}
  \put(0,2.6){{\psfig{figure=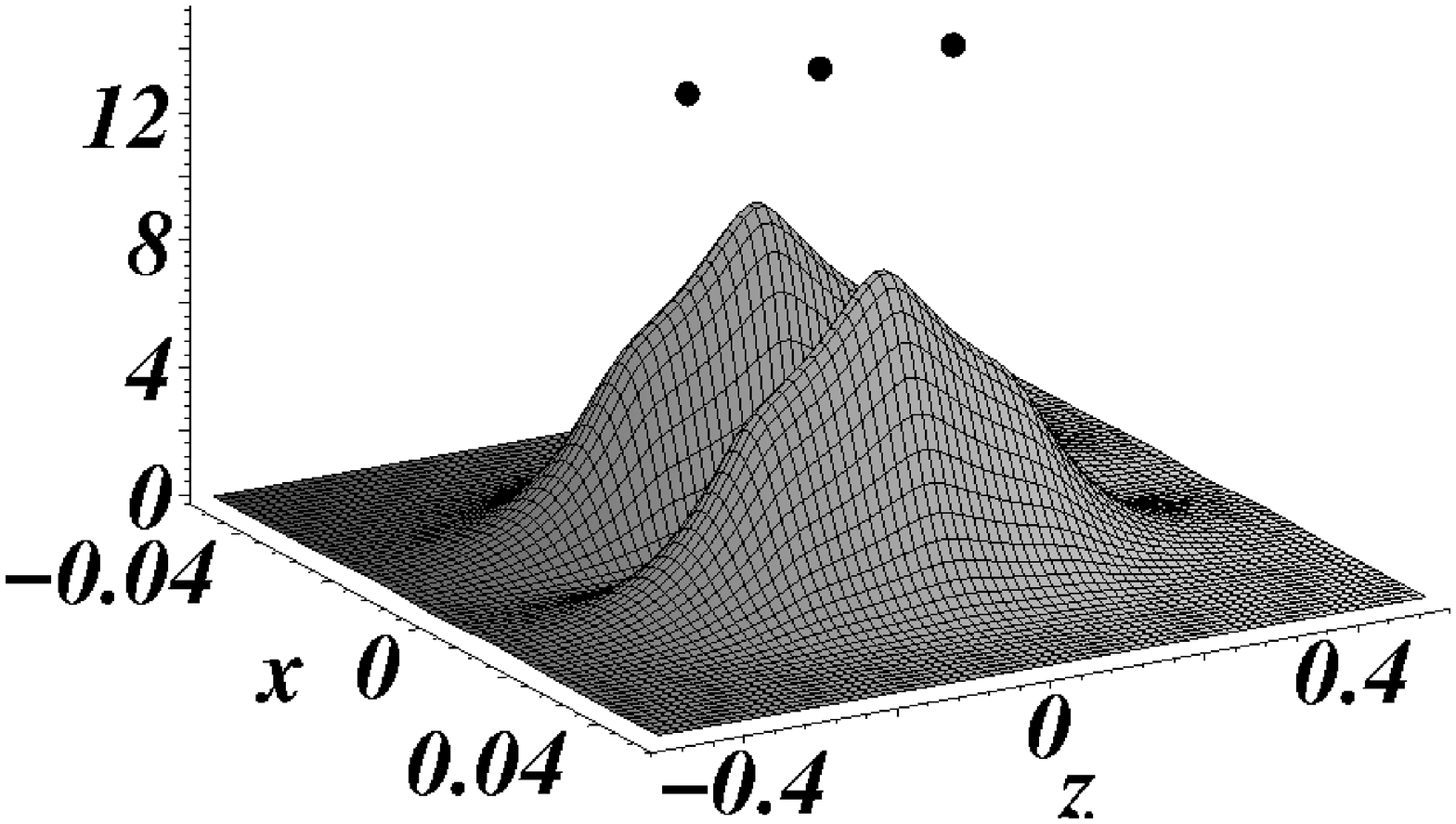,width=3in,angle=-90}}}
  \put(3.6,2.6){{\psfig{figure=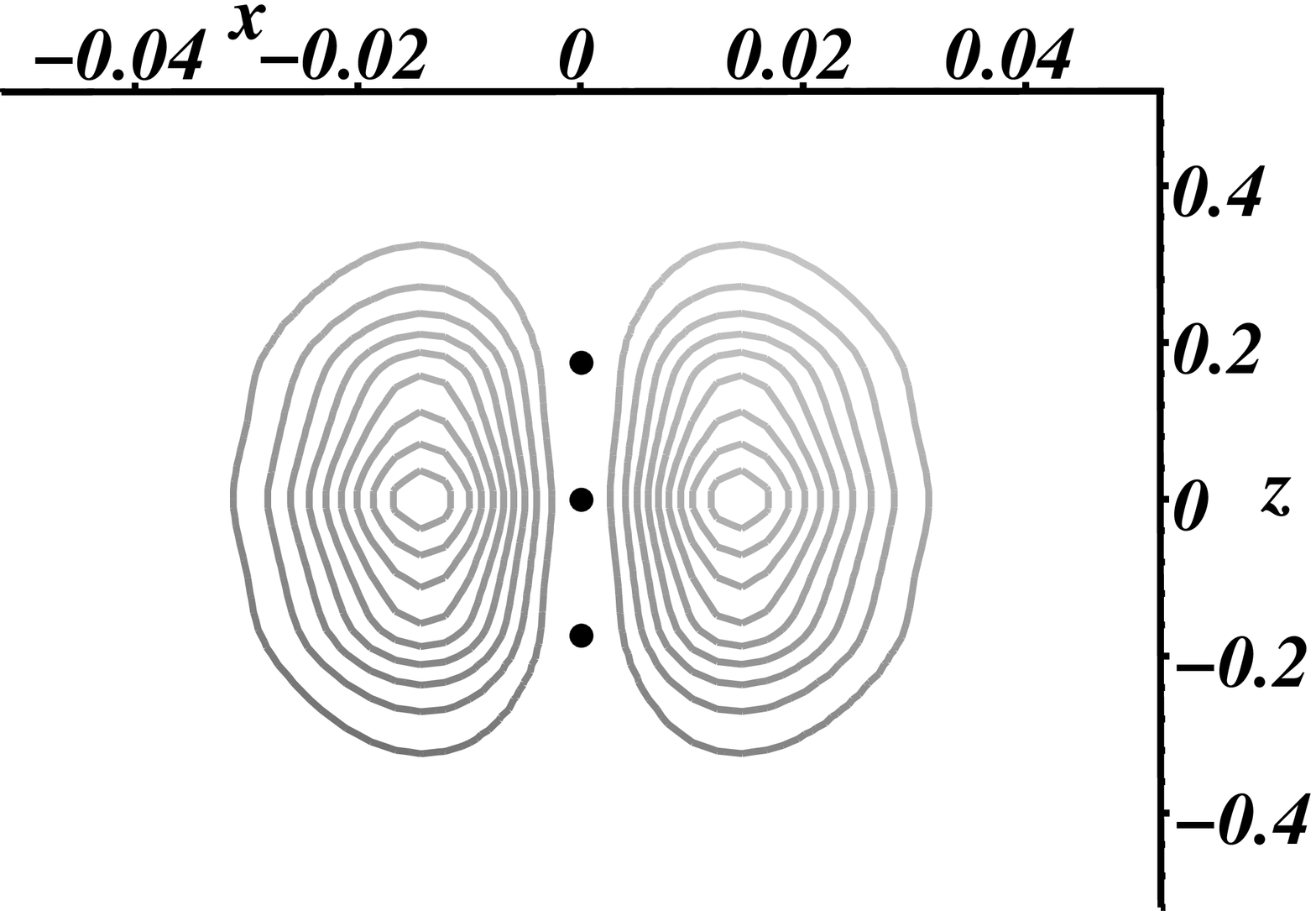,width=3in,angle=-90}}}
%%%%%%%%%%%%%%% 
  \put(0.42,3.9){$|\Psi|^2$} 
  \put(0.42,1.4){$|\Psi|^2$} 
%%%%%%%%%%%%%%%   
  \put(3,3.9){(a)} 
  \put(3,1.4){(b)} 
   \end{picture}
%}
  \end{center}
\caption{\label{fig:edist2}Normalized electronic density distribution
  ${\Psi^2(x,y=0,z)}/{\int \Psi^2(x,y,z) d{\vec r}}$ and the
  corresponding contours for the first two excited states (a) $1\pi_u$
  and (b) $1\delta_g$ of the ion $H_3^{2+}$ in a magnetic field
  $B=10000$\,a.u. The positions of the protons are marked by bullets.
  Vertical axis in the figures on the left is scaled to $1:1000$.}
\end{figure}

\section{Conclusion}

We have carried out an accurate, non-relativistic calculation in the
Born-Oppenheimer approximation for the low-lying states of the
$H_3^{2+}$ molecular ion in the linear parallel configuration placed
in a constant uniform magnetic fields ranging from $B=10^{10}\, G$ to
to $B = 4.414 \times 10^{13}\,G$.  Similar to the study of the $H_2^+$
ion \cite{TurbLopez:2004} we used a variational method with a simple
(and unique), several parametric trial function inspired by the
underlying physics of the problem for all range of magnetic fields
studied.  Our trial function can be easily analyzed and in contrast to
other approaches our results can be easily reproduced and checked.
Also the trial function (\ref{ansatz}) can be easily modified to
explore other excited states.  We showed that the exotic ion
$H_3^{2+}$ exists in the states of positive $z$-parity $1\si_g$,
$1\pi_u$ and $1\de_g$, and does not exist in the states of negative
$z$-parity $1\si_u$, $1\pi_g$ and $1\de_u$ $B=2.35\times 10^{10}\, G -
4.414 \times 10^{13}\,G$.
The present study  complements
our previous study of the ground state performed in
\cite{Turbiner:1999} and \cite{Lopez-Tur:2000}.

It is evident that the state $1\si_g$ having no nodes is the global
ground state of the system $(pppe)$ (if exists) for all magnetic
fields (Perron theorem).  It is clear that this statement remains
valid in general, when even the states other than studied are taken
into account. We show for the $(pppe)$ system in state of positive
$z$-parity the electronic potential surface $E_T(R+,R_-)$ always
develops a minimum corresponding to the symmetric configuration
$R_+=R_-$, when the protons are situated on equal distances between
each other. The domain of existence of the $H_3^{2+}$ ion is slightly
extended in comparison to the previous studies
\cite{Turbiner:1999,Lopez-Tur:2000} to be $B=10^{10} - 4.414 \times
10^{13}\,G$.  For the case of excited states of the positive
$z$-parity ($1\pi_u,1\de_g$) we also find a minimum in the potential
surface for a similar domain of magnetic fields $B=2.35 \times 10^{10}
- 4.414 \times 10^{13}\,G$.  A common feature for these bound states
is that the total and binding energies grow with an increase in the
magnetic field strength, while the internuclear equilibrium distances
reduce drastically.

For fixed magnetic field the energies of the positive $z$-parity
states are ordered following the value of the magnetic quantum number
$m$,
\[
 E_T^{1\si_g} < E_T^{1\pi_u} < E_T^{1\de_g} \ ,
\]
as well as the equilibrium internuclear distances 
\[
 R_{eq}^{1\si_g} < R_{eq}^{1\pi_u} < R_{eq}^{1\de_g} \ .
\]
This order holds also in the whole domain of magnetic fields studied
(see Tables \ref{T1sg}, \ref{T1pu}, and \ref{T1dg}).
The same time for the fixed magnetic field the
binding energies of the states of positive $z$-parity are reduced slow
with excitation (see Tables \ref{T1sg},\ref{T1pu},\ref{T1dg}). It
gives us a chance to expect that other states of the positive
$z$-parity can exist for some magnetic fields.

For the $1\si_g$ state we studied the electronic potential energy
surfaces for different magnetic fields. All those surfaces in addition
to the minimum corresponding to the bound state ($pppe$) display two
symmetric valleys running from the minimum to infinity corresponding
to the ``path'' of the decay $H_3^{2+} \to H_2^+ + p$. For magnetic
fields $B\lesssim 3\times 10^{13}$\,G the $H_3^{2+}$ exotic ion is
unstable towards the decay $H_3^{2+} \to H_2^+ + p$.

%Neglecting
%deviations of the valleys from linearity we give an estimate of the
%lifetime of $H_3^{2+}$, $\Gamma \propto \int V dR$ as a function of
%magnetic field (see Fig.??).

Our analysis of the lowest longitudinal vibrational state and the
height of the potential barrier (see Table~\ref{Tvib}) allow us to
conclude that for magnetic fields $B\gtrsim 2.35\times 10^{12}$\,G,
the well of the potential energy surface of the ground state $1\si_g$
contains at least one vibrational state.

Since for magnetic fields $B\gtrsim 3\times 10^{13}$\,G the $H_3^{2+}$
ion is the most stable one-electron system, this study can be of
considerable importance, in particular, in the construction of
adequate atomic-molecular models of the neutron star atmospheres,
where typical magnetic fields are $B\simeq 10^{12}$\,G or higher.  A
recent application of the present approach, although for larger
magnetic fields where relativistic corrections start to be of a
certain importance (for a discussion see e.g. \cite{Salpeter:1992}),
were used to construct a model of the atmosphere of the isolated
neutron star 1E1207.4-5209 (see \cite{Sandal:2002}) which is based on
the assumption that the main component of such atmosphere is nothing
but the exotic molecular ion $H_3^{2+}$ \cite{NSmodel}.

\begin{acknowledgments}
  This work was supported in part by
  %DGAPA Grant \# IN124202 and by
  CONACyT grants {\bf 25427-E} and {\bf 36600-E} (Mexico). The authors
  are grateful to H. Olivares for a help in numerical calculations.
\end{acknowledgments}

\end{document}